\documentclass[11pt,reqno,twoside]{article}

\usepackage{pgfplots}
\pgfplotsset{compat=1.15}
\usepackage{mathrsfs}
\usetikzlibrary{arrows}

\usepackage[T1]{fontenc}

\usepackage[]{amsmath, amssymb, amsthm, tabularx}

\usepackage[]{a4, xcolor, here}

\usepackage[]{pstricks, pst-text, pst-node, pst-tree, gastex}

\usepackage{tikz}

\usepackage[tikz]{bclogo}

\usepackage{comment}

\usepackage{boites,graphicx}

\usepackage{soul}

\definecolor{pastelyellow}{rgb}{0.99, 0.99, 0.59}
\definecolor{aqua}{rgb}{0.0, 1.0, 1.0} 
\definecolor{aquamarine}{rgb}{0.5, 1.0, 0.83} 
\definecolor{bananayellow}{rgb}{1.0, 0.88, 0.21}
\definecolor{burgundy}{rgb}{0.5, 0.0, 0.13}
\definecolor{ao(english)}{rgb}{0.0, 0.5, 0.0}

\setul{}{0.2ex}
\setulcolor{bananayellow}

\usepackage[normalem]{ulem}

\usepackage{mathrsfs}

\usepackage{stmaryrd}

\usepackage{enumerate}

\usepackage{paralist}

\usepackage{multienum}

\usepackage{multicol}

\usepackage{fancyhdr}

\usepackage{datetime}

\usepackage{everypage}
\usepackage[contents={},opacity=1,scale=1.6,
color=gray!90]{background}
\usepackage{ifthen}

\usepackage[]{hyperref} 

\hypersetup{
	colorlinks = true,
	linkcolor = red,
	anchorcolor = black,
	citecolor = blue, 
	filecolor = cyan,
	menucolor = red,
	runcolor = cyan,
	urlcolor = blue,
	linkbordercolor = {white},
	linktocpage = true
}

\newtheorem{theorem}{Theorem}[section]
\newtheorem{proposition}[theorem]{Proposition}
\newtheorem{lemma}[theorem]{Lemma}
\newtheorem{corollary}[theorem]{Corollary}

\theoremstyle{definition}
\newtheorem{definition}[theorem]{Definition}

\newtheorem{example}[theorem]{Example}
\newtheorem{remark}[theorem]{Remark}

\newcommand{\cC}{\mathcal{C}}

\newcommand{\cF}{\mathcal{F}}
\newcommand{\cG}{\mathcal{G}}

\newcommand{\cP}{\mathcal{P}}

\newcommand{\cS}{\mathcal{S}}
\newcommand{\cU}{\mathcal{U}}
\newcommand{\cV}{\mathcal{V}}

\newcommand{\rsp}[1]{{\mathrm{rowsp}{#1}}}
\newcommand{\lcm}{\mathrm{lcm}}
\newcommand{\stab}{\mathrm{Stab}}
\newcommand{\stabbeta}{\mathrm{Stab}_{\beta}}
\newcommand{\stabsf}{\mathrm{Stab}^+}
\newcommand{\stabsfbeta}{\mathrm{Stab}^{+}_{\beta}}
\newcommand{\orb}{\mathrm{Orb}}
\newcommand{\orbbeta}{\mathrm{Orb}_{\beta}}
\newcommand{\GL}{\mathrm{GL}}

\newcommand{\bbF}{{\mathbb F}} 

\renewcommand{\geq}{\geqslant}
\renewcommand{\leq}{\leqslant}

\begin{document}

	\renewcommand{\headrulewidth}{0pt}
	
	\rhead{ }
	\chead{\scriptsize  Cyclic Orbit Flag Codes}
	\lhead{ }

	\title{Cyclic Orbit Flag Codes
		\renewcommand\thefootnote{\arabic{footnote}}
	}

	\author{\renewcommand\thefootnote{\arabic{footnote}}
		Clementa Alonso-Gonz\'alez\footnotemark[1],\,  Miguel \'Angel Navarro-P\'erez\footnotemark[1]}

	\footnotetext[1]{Dpt.\ de Matem\`atiques, Universitat d'Alacant, Sant Vicent del Raspeig, Ap.\ Correus 99, E -- 03080 Alacant. \\ E-mail adresses: \texttt{clementa.alonso@ua.es, miguelangel.np@ua.es}.}

	{\small \date{\usdate{\today}}} 
	
	\maketitle
	
	\begin{abstract}
		In network coding, a flag code is a set of sequences of nested subspaces of $\bbF_q^n$, being $\bbF_q$ the finite field with $q$ elements. Flag codes defined as orbits of a cyclic subgroup of the general linear group acting on flags of $\bbF_q^n$ are called {\em cyclic orbit flag codes}. Inspired by the ideas in \cite{GLMoTro2015}, we determine the cardinality of a cyclic orbit flag code and provide bounds for its distance with the help of the largest subfield over which all the subspaces of a flag are vector spaces (the {\em best friend} of the flag). Special attention is paid to two specific families of cyclic orbit flag codes attaining the extreme possible values of the distance: \textit{Galois cyclic orbit flag codes} and \textit{optimum distance cyclic orbit flag codes}. We study in detail both classes of codes and analyze the parameters of the respective subcodes that still have a cyclic orbital structure.
	\end{abstract}
	
	

	\section{Introduction}\label{sec:Introduction}
	
	Network coding is a strong tool for effective data transmission in a network modelled as a directed acyclic multigraph with several sources and sinks. In \cite{AhlsCai00}, it was proved that the information flow of the network may be improved if the intermediate nodes are able to perform random linear combinations of the received inputs instead of simply routing them. Random network coding was introduced in \cite{HoMeKo06}, and an algebraic approach to it was presented in \cite{KoetKschi08}. In that work, the authors propose transmitting information by using vector subspaces of $\bbF_q^n$ and define \emph{subspace codes} as a class of codes well suited for error correction. In case all the codewords in a subspace code have the same dimension, it is said to be a \emph{constant dimension code}. The seminal paper \cite{KoetKschi08} has lately led to many lines of research on subspace codes addressed either to the construction of subspace codes with the best size fixed the minimum distance or to find algebraic constructions of subspace codes with good parameters (see \cite{TrautRosen18} and references therein).
	
	In \cite{TrautManRos2010},  Trautmann \emph{et al.} introduced the concept of \emph{orbit codes} as subspace codes obtained from the action of subgroups of the general linear group $\GL(n,q)$ on the set of subspaces of $\bbF_q^n$. When the acting group is cyclic, we speak about \emph{cyclic orbit codes}. This family of codes has awaken a lot of interest due to the simplicity of their algebraic structure and to the existence of efficient encoding/decoding algorithms. We refer the reader to \cite{BenEtGaRa16, ChenLi18, EtVar11, GLLe2021, GLMoTro2015, OtOz17, RothRaTa18, TrautManBraunRos2013, TrautManRos2010, ZhaoTang2019} for some of the more recent papers. 
	
	Taking into account that $\bbF_q^n$ and the field extension $\bbF_{q^n}$ are isomorphic as $\bbF_q$-vector spaces, in \cite{GLMoTro2015}, the authors consider subspace codes as collections of $\bbF_q$-vector subspaces of $\bbF_{q^n}$ and study orbit codes arising from the natural action of the multiplicative subgroups of $\bbF_{q^n}^\ast$ (cyclic groups as well) on $\bbF_q$-vector spaces. Fixed a generating subspace $\cU$ of the cyclic orbit code  $\orb(\cU)$, their main tool is the \emph{best friend} of $\cU$, that is, the largest subfield of $\bbF_{q^n}$ over which $\cU$ is a vector space. This concept is closely related with the stabilizer of $\cU$, specially when the acting group is $\bbF_{q^n}^\ast$. The best friend allows the authors to give relevant information about the cardinality, distance and other features of cyclic orbit codes.

	\emph{Flag codes} were introduced in \cite{LiebNebeVaz18} as a generalization of constant dimension codes in network coding. In a flag code of constant type, codewords are given by sequences of nested subspaces (flags) of prescribed dimensions. In that paper, the multiplicative action of $\GL(n,q)$ is naturally extended from subspaces to flags and several constructions of \emph{orbit flag codes} are provided. In \cite{CasoPlanar, CasoNoPlanar}, flag codes attaining the maximum possible distance (\emph{optimum distance flag codes}) are characterized and obtained without regard to their possible orbital structure whereas in \cite{OrbitODFC} an orbital construction of them is proposed. 
	
	In this work we follow the approach of Gluesing-Luerssen \emph{et al.} in \cite{GLMoTro2015}. Inspired by their ideas, we consider flags on $\bbF_{q^n}$ given by nested $\bbF_q$-subspaces of the field $\bbF_{q^n}$ and focus on  \emph{cyclic orbit flag codes} constructed as orbits of subgroups of $\bbF_{q^n}^\ast$. We  generalize the concept of the best friend of a subspace to the flags framework by defining the \emph{best friend} of a flag as the largest subfield of $\bbF_{q^m}$ over which every subspace in a flag is a vector space. As it occurs in the constant dimension codes scenario, the knowledge of the best friend of a generating flag allows us to easily determine the size of the cyclic orbit code as well as to give estimates for its distance. In particular, we pay special attention to two specific families of cyclic orbit flag codes attaining the extreme possible values of the distance. We introduce first the concept of \emph{Galois cyclic flag codes} as the cyclic orbit codes generated by sequences of nested subfields of $\bbF_{q^n}$. Despite the fact that these codes have the minimum possible distance (fixed the best friend), they present a nice gear of nested spreads compatible with the action of $\bbF_{q^n}^*$. Moreover, if one consider the subcodes of Galois cyclic flag codes that keep cyclic orbital structure, we can improve their distance in a controlled manner and reach even the maximum possible one. By the way, we also determine which dimensions in the type vector of a general generating flag are compatible with attaining the maximum distance, having a fixed best friend and being orbits under the action of subgroups of $\bbF_{q^n}^\ast$. In other words, we study \emph{optimum distance cyclic orbit flag codes} and their orbital cyclic subcodes.
	
	The text is organized as follows. In Section \ref{sec:Preliminaries}, the reader can find the general background on subspace codes. Particular care is devoted to the study of cyclic orbit (subspace) codes developed in \cite{GLMoTro2015}. In Section \ref{sec: cyclic orbit flag codes}, cyclic orbit flag codes are introduced. We also generalize the notions of stabilizer subfield and best friend to the flag codes setting by exhibiting the relationship between these two concepts and the corresponding ones for subspace codes. In Section \ref{sec: prescribed BF}, the cardinality and bounds for the distance of a cyclic orbit flag code with a given best friend are provided. We finish by introducing Galois cyclic flag codes and optimum distance cyclic flag codes with a prescribed best friend. We study their parameters and properties as well as the ones of respective subcodes coming also from the action of subgroups of $\bbF_{q^n}^*$.


	\section{Preliminaries}\label{sec:Preliminaries}
	
	Fix $\bbF_q$ the finite field of $q$ elements where $q$ is a primer power. For any natural number $n\geq 1$, $\bbF_q^n$ represents the $n$-dimensional vector space over $\bbF_q$. Given $1\leq k <n$, the \emph{Grassmannian} $\cG_q(k, n)$ is the set of $k$-dimensional subspaces of $\bbF_q^n$ and we write $\cP_q(n)$ to denote the \emph{projective geometry} of $\bbF_q^n$, that is, the set of all the subspaces of $\bbF_q^n$. 
	The set $\cP_q(n)$ can be considered as a metric space with the \emph{subspace distance} (see \cite{KoetKschi08}) defined as 
	\begin{equation}\label{def: subspace distance}
		d_S(\cU, \cV)= \dim(\cU+\cV)-\dim(\cU\cap\cV).
	\end{equation}
	A \emph{subspace code}  $\cC$ of length $n$ is a nonempty subset of $\cP_q(n)$ and its \emph{minimum subspace distance} is defined as 
	$$
	d_S(\cC)=\min\{ d_S(\cU, \cV) \ | \ \cU, \cV \in \cC, \ \cU \neq \cV \}.
	$$

	A subspace code in which every codeword has the same dimension, say $k$, is called \emph{constant dimension code} of dimension $k$ and length $n$ (see \cite{TrautRosen18} and references therein). The subspace distance between two subspaces $\cU$ and $\cV$ of dimension $k$ is given by
$$
		d_S(\cU, \cV)= 2(k-\dim(\cU\cap\cV)).
$$
Consequently, the minimum distance of a constant dimension code of dimension $k$ is upper bounded by
	\begin{equation}\label{eq: bound subspace distance}
		d_S(\cC)\leq
		\left\lbrace
		\begin{array}{lll}
			2k      & \text{if} & 2k\leq n, \\
			2(n-k)  & \text{if} & 2k > n.
		\end{array}
		\right.
	\end{equation}
	These bounds for the distance are attained by constant dimension codes in which every pair of codewords intersects with the minimum possible dimension. For dimensions $k$ up to $\lfloor\frac{n}{2}\rfloor$, constant dimension codes attaining the previous bound are known as \emph{partial spread codes} and their cardinality is, at most, $\lfloor \frac{q^n-1}{q^k-1}\rfloor.$ 
	In contrast, a constant dimension code that attains the bound in (\ref{eq: bound subspace distance}) and has dimension $k > \lfloor \frac{n}{2}\rfloor$, cannot contain more than $\lfloor \frac{q^n-1}{q^{n-k}-1}\rfloor$ elements. 
	
	A \emph{spread code} in $\cG_q(k, n)$, or just a \emph{$k$-spread}, is a partition of $\bbF_q^n$ into $k$-dimensional subspaces. In other words, a spread is a partial spread that covers $\bbF_q^n$. Spreads are classical objects coming from Finite Geometry and it is well known that $k$-spreads exist if, and only if, $k$ divides $n$ (see \cite{Segre64}). As a consequence, the size of  every $k$-spread is $\frac{q^n-1}{q^k-1}.$ For further information related to spread codes in the network coding framework, we refer the reader to \cite{GoManRo12, MangaGorlaRosen08, MangaTraut14, TrautRosen18}.

	There are constant dimension codes that can be obtained as orbits of the action of subgroups of the general linear group $\GL(n, q)$ on the Grassmannian of the corresponding dimension. In this case, we speak about \emph{orbit codes}, which were introduced for the first time in \cite{TrautManRos2010}. Given a $k$-dimensional  subspace $\cU$ of $\bbF_q^n$ and a subgroup $G$ of $\GL(n, q)$, the orbit of $\cU$ under the action of $G$ is the constant dimension code given by $\mathrm{Orb}_G(\cU) = \{ \cU \cdot A \ | \ A\in G\},$ where $\cU\cdot A = \rsp (UA)$ for any full-rank generator matrix $U$ of $\cU$. The \emph{stabilizer} of $\cU$ under the action of $G$ is the subgroup $\mathrm{Stab}_G(\cU)= \{ A\in G  \ | \ \cU\cdot A= \cU \}.$ Clearly,
	$$
		| \mathrm{Orb}_G(\cU)| = \frac{|G|}{|\mathrm{Stab}_G(\cU)|}
	$$
	and its minimum distance  is given by
	$$
	d_S( \mathrm{Orb}_G(\cU))= \min \{ d_s(\cU, \cU\cdot A) \ | \ A \in G \setminus\mathrm{Stab}_G(\cU) \}.
	$$
	If the group $G$ is cyclic, the code $\mathrm{Orb}_G(\cU)$ is called \emph{cyclic orbit code.} This special family of orbit codes was widely studied in \cite{GLMoTro2015, ManTrautRos11, RosTraut2013, TrautManBraunRos2013}. In particular, using the fact that $\bbF_q^n$ and $\bbF_{q^n}$ are isomorphic as $\bbF_q$-vector spaces, Trautmann \textit{et al.} provide in  \cite{TrautManBraunRos2013} the following construction of a $k$-spread as a cyclic orbit code. Take a divisor $k$ of $n$ and let $\alpha$ denote a primitive element of  $\bbF_{q^n}$, i.e, a generator of the multiplicative group $\bbF_{q^n}^\ast.$ If we put $c=\frac{q^n-1}{q^k-1},$ then it is clear that $\langle \alpha^c\rangle$ is the unique subgroup of order $q^k-1$ of $\bbF_{q^n}^\ast$ and that $\langle\alpha^c\rangle\cup\{0\}=\bbF_{q^k}.$ As proved in \cite[Th. 31]{TrautManBraunRos2013}, the stabilizer of $\bbF_{q^k}$ under the action of the cyclic group $\langle\alpha\rangle$ is precisely  the subgroup $\langle\alpha^c\rangle$ and the orbit
	\begin{equation}\label{def: spread Anna-Lena}
		\cS = \orb_{\langle \alpha \rangle}(\bbF_{q^k})= \{ \bbF_{q^k}\alpha^i \ | \ i=0, \ldots, c-1 \}
	\end{equation}
	is a $k$-spread of $\bbF_{q^n}$.

	In \cite{GLMoTro2015}, Gluesing-Luerssen \textit{et al.} generalize the construction in (\ref{def: spread Anna-Lena}) for any $\beta \in \mathbb{F}_{q^n}^*$ by introducing the concept of $\beta$-\emph{cyclic orbit code generated by a subspace $\cU$} of $\mathbb{F}_{q^n}$ and study these codes by specifying the largest subfield over which the subspace $\cU$ is a vector space. Let us recall some definitions and results from that work that we will use along this paper.
	
	Consider any nonzero element $\beta$ in the finite field $\bbF_{q^n}$ and the natural multiplicative action of the group $\langle \beta \rangle$ on $\bbF_q$-vector subspaces of $\bbF_{q^n}.$ Orbits of this action are called $\beta$-\emph{cyclic orbit codes}. To be precise, if $1\leq k<n$ and $\cU\subset \bbF_{q^n}$ is a $k$-dimensional subspace over $\bbF_q$, the $\beta$-\emph{cyclic orbit code generated by $\cU$} is the constant dimension code in the Grassmannian $\cG_q(k, n)$ given by
	$$
	\orbbeta(\cU)= \{\cU \beta^i \ | \ 0\leq i \leq |\beta|-1\},
	$$
	where $|\beta|$ denotes the \emph{multiplicative order} of $\beta$ (for further information on these orbits, see \cite{Drudge2002}). The \emph{stabilizer} of the subspace $\cU$ under the action of $\langle \beta \rangle$ is the  cyclic subgroup defined as $\stabbeta(\cU)= \{\beta^i \in \langle\beta\rangle \ | \ \cU\beta^i= \cU \}$ and the \emph{stabilizer subfield}  $\stabsfbeta(\cU)$ of $\cU$ (with respect to $\beta$) is the smallest subfield of $\bbF_{q^n}$ containing both $\bbF_q$ and $\stabbeta(\cU).$
	
	\begin{remark}
		When the acting group is $\bbF_{q^n}^\ast$, following the notation in \cite{GLMoTro2015}, we simply write by $\orb(\cU)$ and call it the \emph{cyclic orbit code generated by $\cU$}. In this situation, we also remove the subscript $\beta$ and write $\stab(\cU)$ and $\stabsf(\cU)$ to the denote the stabilizer and the stabilizer subfield of $\cU$ respectively.
	\end{remark}
	
	Concerning the cardinality of a $\beta$-cyclic orbit code, there exists a nice relationship between $|\orbbeta(\cU)|$ and the dimension of the generating subspace $\cU$. More precisely,  in \cite[Prop. 3.7]{GLMoTro2015},  the authors showed that, if $\cU$ is a $k$-dimensional subspace of $\mathbb{F}_{q^n}$, then
	\begin{equation}\label{eq:cardinal}
		|\beta^{q^k-1}|=\frac{|\beta|}{\gcd(|\beta|,q^k-1)} \textrm{  divides  } | \orbbeta(\cU)|.
	\end{equation}
	Moreover, the equality $| \orbbeta(\cU)|= \frac{|\beta|}{q^k-1}$ holds if, and only if, $\cU$ is a vector space over $\bbF_{q^k}$. More precisely, if $1\in\cU$, for every divisor $k$ of $n$, the code $\orb(\cU)$ is a $k$-spread if, and only if, $\cU=\bbF_{q^k}$. Therefore, the spread defined in (\ref{def: spread Anna-Lena}) arises as the cyclic orbit code $\orb(\bbF_{q^k})$ in this context. 
	
	A subfield $\bbF_{q^m}$ of $\bbF_{q^n}$ is said to be a \emph{friend} of a subspace $\cU \subset \bbF_{q^n} $ if $\cU$ is an $\bbF_{q^m}$-vector space. In that case, if $t$ is the dimension of $\cU$ as $\bbF_{q^m}$-vector space, we have that $\dim_{\bbF_q}(\cU) = mt$. 
	Moreover, if $\{u_1, \ldots, u_t\}\subseteq \cU$ is a basis of $\cU$ over $\bbF_{q^m},$ then it holds 
	$$
		\cU = \bbF_{q^m}u_1 \oplus \cdots \oplus  \bbF_{q^m}u_t.
	$$
Note that every subspace $\cU$ is a vector space over $\stabsfbeta(\cU)$. In other words, the stabilizer subfield is a friend of $\cU$. The largest friend of $\cU$ is called its \emph{best friend} (see  \cite{GLMoTro2015}). The concepts of stabilizer subfield and best friend of a subspace turn to be same in the following situation in which, in addition, the knowledge of the best friend of $\cU$ provides straightforwardly the cardinality of the cyclic orbit code as well as a lower bound for its distance.

	\begin{proposition}\label{prop: stab+ es best friend}(\cite[ Prop. 3.3, 3.12, 3.13 and 4.1]{GLMoTro2015})
		If $\cU$ is a subspace of $\bbF_{q^n}$, then its stabilizer subfield satisfies 
		$$
		\stabsf(\cU)= \stab(\cU) \cup \{0 \}
		$$
		and it contains every friend of $\cU$. As a consequence, the field $\stabsf(\cU)$ is the best friend of the subspace $\cU$. In particular, if $\stabsf(\cU)=\bbF_{q^m}$, then 
		$$
		|\orb(\cU)|=\frac{q^n-1}{q^m-1}.
		$$
		Moreover, the value $2m$ divides the distance between every pair of subspaces in $\orb(\cU)$ and, hence, we have that  $d_S(\orb(\cU))\geq 2m.$ Besides, if $1\in \cU$, we have the inclusion $\stabsf(\cU)\subseteq \cU$.
	\end{proposition}

	\section{Cyclic orbit flag codes}\label{sec: cyclic orbit flag codes}
	In classical linear algebra, a flag variety on the field extension $\bbF_{q^n}$ is a homogeneous space that generalizes the Grassmann variety and whose points are flags. The use of flags in network coding was proposed for the first time in \cite{LiebNebeVaz18}.  We start this section by recalling some basic background on flag codes. Next, we will focus on the family of flag codes that are orbits under the action of a cyclic group on the flag variety. Finally, we introduce the concepts of stabilizer subfield and best friend of a flag, following the ideas in \cite{GLMoTro2015}, in order to deepen the structure and properties of the family of cyclic orbit flag codes.
	
	\subsection{Flag codes}
	\begin{definition}
		A {\em flag} $\mathcal{F}=(\mathcal{F}_1,\ldots,  \mathcal{F}_r)$ on $\mathbb{F}_{q^n}$ is a sequence of nested $\bbF_q$-vector subspaces of $\mathbb{F}_{q^n}$, i.e., such that  
		$$
		\{0\}\subsetneq \mathcal{F}_1 \subsetneq \cdots \subsetneq \mathcal{F}_r \subsetneq \mathbb{F}_q^n.
		$$
		The subspace $\mathcal{F}_i$ is said to be the {\em $i$-th subspace} of $\cF$. The type of $\mathcal{F}$ is the vector $(\dim(\cF_1), \dots, \dim(\cF_r))$. In case the type vector is $(1, 2, \ldots, n-1),$ we say that ${\cF}$ is a {\em full flag}.
	\end{definition}

	The \emph{flag variety} $\mathcal{F}_q((t_1,\dots,t_r),n)$ is the set of flags of type $(t_1, \dots, t_r)$ on $\mathbb{F}_{q^n}$. This variety can naturally be equipped with a metric by extending the subspace distance defined in (\ref{def: subspace distance}). Given two flags $\cF=(\mathcal{F}_1,\ldots,  \mathcal{F}_r)$ and $\cF'=(\mathcal{F}'_1,\ldots,  \mathcal{F}'_r)$ in $\mathcal{F}_q( (t_1, \ldots, t_r),n)$, their \emph{flag distance} is
	$$
	d_f(\cF,\cF')= \sum_{i=1}^r d_S(\mathcal{F}_i, \mathcal{F}'_i).
	$$
	\begin{definition}
		A \emph{flag code} of type $(t_1, \dots, t_r)$ on $\bbF_{q^n}$ is a nonempty subset $\cC\subseteq \cF_q((t_1, \dots, t_r), n)$. Its {\em minimum distance} is given by
		$$
		d_f(\cC)=\min\{d_f(\cF,\cF')\ |\ \cF,\cF'\in\cC, \ \cF\neq \cF'\}
		$$
		and, in case $|\cC|=1$, we put $d_f(\cC)=0.$
	\end{definition}

For each dimension $t_i$ in the type vector of a flag code $\cC$, we can associate to it the constant dimension code in the Grassmannian $\cG_q(t_i, n)$ consisting of the set of the $i$-th subspaces of flags in $\cC$. This set is called the \emph{$i$-projected code} of $\cC$ and we denote it by $\cC_i$. It is clear that  $\vert \cC_i\vert \leq \vert \cC \vert$ for every $i=1, \dots, r$. In case  $|\cC_1|=\dots=|\cC_r|=|\cC|$, we say that  $\cC$ is \emph{disjoint}. As shown in  \cite{CasoPlanar}, the property of being disjoint is necessary in order to have flag codes that achieve the maximum possible flag distance. For type $(t_1, \dots, t_r),$ that maximum distance is
	\begin{equation}\label{eq: dist max flags}
		2 \left( \sum_{t_i \leq \lfloor \frac{n}{2}\rfloor} t_i + \sum_{t_i > \lfloor \frac{t}{2}\rfloor} (n-t_i) \right)
	\end{equation}
and flag codes attaining it are called \emph{optimum distance flag codes}. In \cite{CasoPlanar, CasoNoPlanar} the reader can find constructions of this class of codes as well as the following characterization of them.
	
	\begin{theorem}\cite[Th. 3.11]{CasoPlanar}\label{theo: caracterización ODFC}
		A flag code is an optimum distance flag code if, and only if, it is disjoint and every projected code attains the maximum possible distance for its dimension.
	\end{theorem}

	As in the case of subspace codes, one can build families of flag codes through the action of a group. This approach already appears in \cite{LiebNebeVaz18}, where the authors generalize the action of $\GL(n, q)$ on subspaces of $\bbF_q^n$ to flags and provide several constructions of flag codes as orbits of the action of specific upper unitriangular matrix groups on the full flag variety.

	In the next section, following the ideas developed in \cite{GLMoTro2015} for subspace codes, we introduce the concept of \emph{cyclic orbit flag code} as the orbit of the multiplicative action of subgroups of $\bbF_{q^n}^\ast$ on flags on $\bbF_{q^n}$. 
	
	\subsection{Cyclic orbit flag codes}
	
	Given a nonzero element $\beta$ in the field $\bbF_{q^n}$, we can extend the natural action of the cyclic group $\langle \beta \rangle$ on $\bbF_q$-subspaces of $\bbF_{q^n}$ to flags on $\bbF_{q^n}$ as follows.  If $\cF=(\mathcal{F}_1,\dots,  \mathcal{F}_r)$ is a flag of type $(t_1, \ldots, t_r)$ on $\bbF_{q^n}$, we define the flag $\cF \beta $ as 
	$$
		\cF\beta =  (\cF_1\beta, \ldots, \cF_r\beta).
	$$
	The set 
	\begin{equation}\label{def: cyclic orbit flag code}
		\orbbeta(\cF) = \{ \cF \beta^j \ | \ 0\leq j \leq |\beta|-1 \}.
	\end{equation}
	is called the $\beta$-\emph{cyclic orbit flag code} generated by $\cF.$ The \emph{stabilizer} of the flag $\cF$ (w.r.t. $\beta$) is the subgroup of $\langle \beta \rangle$ given by
	$$
		\stabbeta(\cF)= \{\beta^j \in \langle\beta\rangle \ | \ \cF\beta^j= \cF \}.
	$$
	When the acting group is $\bbF_{q^n}^\ast$, we do not specify it and simply write $\orb(\cF)$ to denote the \emph{cyclic orbit flag code generated by $\cF$.} We also drop the subscript in $\stab(\cF)$. Observe that every $\orbbeta(\cF)$ is a subcode of $\orb(\cF)$. Furthermore, it holds
			$$
			\stabbeta(\cF)=\langle\beta\rangle \cap \stab(\cF).
			$$
	
	As in the subspace codes framework, the orbital structure simplifies the computation of the code parameters: the cardinality of the flag code in (\ref{def: cyclic orbit flag code}) is given by
	\begin{equation}\label{cardinality cyclic orbit flag code}
		| \orbbeta(\cF)| = \dfrac{|\beta|}{|\stabbeta(\cF)|}  = \dfrac{|\beta|}{|\langle\beta\rangle \cap \stab(\cF)|}
	\end{equation}
	and its minimum distance can be computed as
		$$
		d_f(\orbbeta(\cF))= \min\{ d_f(\cF, \cF\beta^j) \ | \ \beta^j \notin \stabbeta(\cF) \}.
		$$
	\begin{remark}
		Notice that the projected codes associated to  $\orbbeta(\cF)$ are $\beta$-cyclic orbit (subspace) codes as well. More precisely, for every $1\leq i \leq r$, we have
			$$
			(\orbbeta(\cF))_i = \orbbeta(\cF_i).
			$$		
		Moreover, as for any other group action, it holds a clear relationship between the stabilizer of the flag $\cF$ and the ones of its subspaces:
		\begin{equation}\label{eq: estabilizador flag}
			\stabbeta(\cF)=\bigcap_{i=1}^r \stabbeta(\cF_i). 
		\end{equation}
	\end{remark}
	
	This equality leads to a direct link between the cardinality of a $\beta$-cyclic orbit flag code, the ones of its projected codes, and the dimensions on the generating flag type vector.
	
	\begin{proposition}\label{prop: cardinal proyectado divide}
		Let $\cF=(\cF_1, \ldots, \cF_r)$ be a flag of type $(t_1, \ldots, t_r)$ on $\bbF_{q^n}$ and  $\beta \in \bbF_{q^n}^*.$ Then $|\orbbeta(\cF_i)|$ divides $|\orbbeta(\cF)|$, for $1\leq i\leq r$. In particular, 
		$$
		\lcm\left\lbrace |\beta^{q^{t_i}-1}| \ | \ 1\leq i \leq r \right\rbrace \ \text{divides} \ |\orbbeta(\cF)|.
		$$
	\end{proposition}
	
	\begin{proof}
		Recall that $|\orbbeta(\cF)| = \frac{|\beta|}{|\stabbeta(\cF)|}$ and $|\orbbeta(\cF_i)| = \frac{|\beta|}{|\stabbeta(\cF_i)|}$, for eve\-ry $1 \leq i\leq r$. Moreover, by means of (\ref{eq: estabilizador flag}), we have that $|\stabbeta(\cF)|$ divides $|\stabbeta(\cF_i)|$ for every value of $i$. Hence, the cardinality of $\orbbeta(\cF_i)$ must divide $|\orbbeta(\cF)|,$  for $1 \leq i\leq r$. The last part of the statement follows directly from this fact along with (\ref{eq:cardinal}).
	\end{proof}

	\subsection{Stabilizer subfield and best friend of a flag code}
	The following definition extends the concept of stabilizer subfield of a subspace defined in \cite{GLMoTro2015} to the flag codes setting.
	
	\begin{definition}
		Let $\cF=(\cF_1, \ldots, \cF_r)$ be a flag of type $(t_1, \ldots, t_r)$ on the field $\bbF_{q^n}$ and $\beta\in\bbF_{q^n}^\ast$. We define the \emph{stabilizer subfield} of the flag $\cF$ (w.r.t. $\beta$) as the smallest subfield $\stabsfbeta(\cF)$ of $\bbF_{q^n}$ containing both $\bbF_q$ and $\stabbeta(\cF)$.
	\end{definition}
	
	As before, if $\beta$ is a primitive element of $\bbF_{q^n}$, we just write $\stabsf(\cF)$. In this case, the stabilizer subfield of a flag admits the following nice description:
	
	\begin{proposition}\label{prop: stab+ flag}
		Let $\cF=(\cF_1, \ldots, \cF_r)$ be a flag on $\bbF_{q^n}$. It holds
		$$
		\stabsf(\cF)= \stab(\cF)\cup \{0\}=\bigcap_{i=1}^r\stabsf(\cF_i)
		$$
		and every $i$-th subspace $\cF_i$ of the flag $\cF$ is a vector space over $\stabsf(\cF)$. Moreover, if $1\in\cF_1$, the stabilizer subfield $\stabsf(\cF)$ is contained in every subspace of $\cF$.
		
	\end{proposition}
	\begin{proof}
		By application of Proposition \ref{prop: stab+ es best friend}, one has that $\stabsf(\cF_i)=\stab(\cF_i)\cup\{0\}$ for every $1\leq i\leq r$. Now, by means of (\ref{eq: estabilizador flag}), we conclude that
		
		\begin{equation}\label{stab+ de un flag como intersección}
			\begin{array}{ccccc}
				\stab(\cF) \cup \{0\} & = & \left(\bigcap_{i=1}^r \stab(\cF_i)\right) \cup\{0\}   &   &                              \\
				& = & \left( \bigcap_{i=1}^r \stab(\cF_i) \cup\{0\}  \right)& = & \bigcap_{i=1}^r\stabsf(\cF_i).
			\end{array}
		\end{equation}
		This proves that $\stab(\cF) \cup \{0\}$ is a field and then it is the stabilizer subfield of the flag $\cF$. Moreover, it is a subfield of every $\stabsf(\cF_i)$. Hence, it is clear that the subspace $\cF_i$ is a vector space over $\stabsf(\cF)$. Besides, if $1\in \cF_1$, by using Proposition \ref{prop: stab+ es best friend}, we obtain
		$$
		\stabsf(\cF)\subseteq \stabsf(\cF_1)\subseteq \cF_1 \subset \cF_2 \subset \dots \subset \cF_r.
		$$
	\end{proof}
	
	Notice that the condition $1\in \cF_1$ in Proposition \ref{prop: stab+ flag} is by no means restrictive when the acting group is $\bbF_{q^n}^\ast$. In fact, we can always find a generating flag fulfilling this property. It suffices to see that, given an arbitrary flag $\cF$, for every nonzero element $\beta\in \cF_1$, the flag $\cF\beta^{-1}$ clearly satisfies the required condition. Moreover, since $\beta$ is an element in the field $\bbF_{q^n}^\ast$, both flags $\cF$ and $\cF\beta^{-1}$ generate the same cyclic orbit flag code $\orb(\cF)=\orb(\cF\beta^{-1})$.
	
	\begin{remark}\label{rem: vector space over stabsfbeta}
		Clearly, if $\beta \in \bbF_{q^n}^*,$ it holds $\stabbeta(\cF) \subseteq \stab(\cF)$ and, hence, $ \stabsfbeta(\cF) \subseteq \stabsf(\cF)$. As a consequence, every $\cF_i$ is a vector space over the field $\stabsfbeta(\cF)$ as well as over all its subfields. Moreover, if $1 \in \cF_1,$ then $\stabsfbeta(\cF) \subseteq \cF_i$ for $1\leq i\leq r$.
	\end{remark} 
\noindent As it occurs for constant dimension codes, the inclusion $ \stabsfbeta(\cF) \subseteq \stabsf(\cF)$ may be strict. Let us provide an example from a length-two flag inspired by \cite[Example 3.6]{GLMoTro2015}. 
	
	\begin{example}\label{ex: inclusion estricta stabsfbeta}
		Consider the flag $\cF=(\bbF_{3^2} , \bbF_{3^4})$ on the field $\bbF_{3^8}$ and let $\alpha$ be a primitive element of $\bbF_{3^8}$. Observe that $\stabsf(\bbF_{3^2})=\bbF_{3^2}$ and $\stabsf(\bbF_{3^4})=\bbF_{3^4}$. Hence, by Proposition \ref{prop: stab+ flag}, it follows that $\stabsf(\cF)=\bbF_{3^2}\cap \bbF_{3^4} = \bbF_{3^2}.$ Let us now choose $\beta= \alpha^{1312}$, which have multiplicative order equal to $5$. Observe that $\stabbeta(\cF)\subseteq \langle\beta\rangle$ and also $\stabbeta(\cF)\subseteq \stabsfbeta(\cF)^\ast \subseteq \stabsf(\cF)^\ast = \bbF_{3^2}^\ast$. As the orders of $\langle\beta\rangle$ and $\bbF_{3^2}^\ast$ are coprime, we have that $\stabbeta(\cF)= \{1\}$. This implies that $\stabsfbeta(\cF)=\bbF_3$.
	\end{example}

	There are remarkable connections between the cardinality of a $\beta$-cyclic orbit flag code and the generating flag when one has a divisor of $n$ among the dimensions of the type vector.
	
	\begin{proposition}\label{prop: cardinal beta-cíclico}
		Let $\cF=(\cF_1, \ldots, \cF_r)$ be a flag of type $(t_1, \ldots, t_t)$ on $\bbF_{q^n}$. Assume that $m$ is a divisor of $n$ such that $m=t_i$ for some $i\in \{1,\ldots, r\}$ and consider the subfield $\bbF_{q^m}$ of $\bbF_{q^n}$. Take an element $\beta \in \bbF_{q^n}^\ast$ such that $\bbF_{q^m}^\ast \subseteq \langle\beta\rangle$. Then:
		\begin{enumerate}
			\item The value $\frac{|\beta|}{q^m-1}$ divides $|\orbbeta(\cF)|.$ \label{item1}
			\item We have $|\orbbeta(\cF)|= \frac{|\beta|}{q^m-1}$ if, and only if, each subspace $\cF_j$ is a vector space over $\bbF_{q^m}.$ In particular, $t_1=m$. \label{item2}
		\end{enumerate}
	\end{proposition}
	\begin{proof}
		As $\bbF_{q^m}^\ast\subseteq \langle\beta\rangle$, we have that $q^m-1$ must divide $|\beta|$. This implies that $|\beta^{q^{t_i}-1}|=|\beta^{q^m-1}|=\frac{|\beta|}{q^m-1}$ and (\ref{item1}) follows directly from Proposition \ref{prop: cardinal proyectado divide}.
		
		To prove (\ref{item2}), observe that $|\orbbeta(\cF)|= \frac{|\beta|}{q^m-1}$ holds if, and only if, $\stabbeta(\cF)$ is a subgroup of order $q^m-1$ of $\langle\beta\rangle$. By the uniqueness of subgroups of a cyclic group, it follows that $\stabbeta(\cF)=\bbF_{q^m}^\ast$. Hence, the field $\stabsfbeta(\cF)=\bbF_{q^m}$ is a subfield of $\stabsf(\cF)$ and, by means of Remark \ref{rem: vector space over stabsfbeta}, every subspace $\cF_j$ has structure of $\bbF_{q^m}$-vector space. In particular, no dimension smaller than $m$ can appear in the type vector, i.e., $t_1=m$.
		
		Conversely, assume that every $\cF_j$ is a vector space over $\bbF_{q^m}$ for $j\in \{1, \ldots, r\}$. In particular, $\cF_1=\bbF_{q^m}\gamma$ for some $\gamma\in \bbF_{q^n}^*$. As a consequence, multiplication by elements in $\bbF_{q^m}^\ast\subseteq \langle\beta\rangle$ is closed on every subspace $\cF_j$. Hence, we have $\bbF_{q^m}^\ast \subseteq \stabbeta(\cF_j)$ for $1\leq j\leq r$ and, by means of (\ref{eq: estabilizador flag}), it holds $\bbF_{q^m}^\ast\subseteq \stabbeta(\cF)$. On the other hand, notice that $\stabbeta(\cF)\subseteq \stabbeta(\cF_1) = \bbF_{q^m}^\ast$.  Thus, it follows that $\stabbeta(\cF)=\bbF_{q^m}^\ast$ and $|\orbbeta(\cF)|=\frac{|\beta|}{q^m-1},$ as we wanted to prove.
	\end{proof}
	
	The second statement in Proposition \ref{prop: cardinal beta-cíclico} turns out specially interesting in the case of cyclic orbit codes, that is, when the acting group is $\bbF_{q^n}^\ast$.

	\begin{corollary}\label{cor: cardinal cíclico}
		Let  $\cF=(\cF_1, \ldots, \cF_r)$ be a flag of type $(t_1, \ldots, t_r)$ on $\bbF_{q^n}$. Assume that $m$ is a divisor of $n$ such that $m=t_i$ for some $i\in \{1,\ldots, r\}$. If $|\orb(\cF)|= \frac{q^n-1}{q^m-1}$, then $m=t_1$ and the constant dimension code $\orb(\cF_1)$ is the $m$-spread $\orb(\bbF_{q^m})$. Moreover, the value $m$ divides $t_j$, for $j\in \{1, \ldots, r\}$.
	\end{corollary}
	\begin{proof}
		By means of Proposition \ref{prop: cardinal beta-cíclico}, it is clear that the first dimension in the type vector is $t_1=m$ and it divides every $t_i$. Moreover, $\cF_1$ must be a one-dimensional vector space over $\bbF_{q^m}$, that is, it is of the form $\cF_1=\bbF_{q^m}\gamma$ for some $\gamma\in \bbF_{q^n}^*$. As a result, the first projected code $\orb(\cF_1)=\orb(\bbF_{q^m})$ is the $m$-spread defined in (\ref{def: spread Anna-Lena}).
	\end{proof}
	
	\begin{remark}
		In the conditions of the previous corollary, if we require the subspace $\cF_1$ to contain the element $1\in \bbF_{q^n}$, not only do we obtain that  $\orb(\cF_1)=\orb(\bbF_{q^m})$ but also the equality $\cF_1=\bbF_{q^m}$.
	\end{remark}

	In view of Propositions \ref{prop: stab+ flag} and \ref{prop: cardinal beta-cíclico}, it also makes sense the extension to flags of the concept of best friend introduced in \cite{GLMoTro2015}.  
	\begin{definition}
		Consider a flag $\cF$ on $\bbF_{q^n}$. A subfield $\bbF_{q^m}$ of $\bbF_{q^n}$ is said to be a \emph{friend} of the flag $\cF$ if all its subspaces are $\bbF_{q^m}$-vector spaces. In other words, a subfield of $\bbF_{q^n}$ is a friend of the flag $\cF$  if it is a friend of all its subspaces. We call \emph{best friend} of the flag $\cF$ to its largest friend.
	\end{definition}
	
	The next result states a necessary condition on the type vector of flags having a given subfield of $\bbF_{q^n}$ as a friend. The proof is straightforward.
	\begin{lemma}\label{lem: BF divides dimensions}
		Let $\cF=(\cF_1, \ldots, \cF_r)$ be a flag of type $(t_1, \ldots, t_r)$ on $\bbF_{q^n}$. If $\bbF_{q^m}$ is a friend of $\cF$ then $m$ divides  $\gcd(t_1, \ldots, t_r, n).$
	\end{lemma}
	\begin{remark}
		If follows that the best friend of a flag of type $(t_1, \ldots, t_r)$ with $\gcd(t_1, \ldots, t_r, n)=1$, in particular a full flag, is the ground field $\bbF_q$.
	\end{remark}

	Beyond conditions on the type vector, we can always characterize the best friend of an arbitrary flag in terms of the ones of its subspaces. To do so, we generalize Proposition \ref{prop: stab+ es best friend} to the flag codes scenario. 
	
	\begin{proposition}\label{prop: stab+ es el best friend del flag}
		Let $\cF=(\cF_1, \ldots, \cF_r)$ be a flag on $\bbF_{q^n}$. Then 
		$\stabsf(\cF)$ is the best friend of the flag $\cF$ and it contains any other friend $\bbF_{q^m}$ of $\cF$. Moreover, if $1\in \cF_1$, then we have that $\bbF_{q^m} \subseteq \stabsf(\cF) \subseteq \cF_1$. 
	\end{proposition}
	\begin{proof}
		Let us prove that $\stabsf(\cF)$ is the largest friend of $\cF$, i.e., its best friend. To do so, assume that a subfield $\bbF_{q^m}$ of $\bbF_{q^n}$ is a friend of the flag $\cF$. By definition of friend of a flag, we know that multiplication by elements in $\bbF_{q^m}$ is closed in every subspace $\cF_i$ of the flag. As a consequence, $\bbF_{q^m}^\ast$ is a subgroup of  $\stab(\cF)$ and we can conclude that $\bbF_{q^m}$ is contained in $\stab(\cF)\cup\{0\} =  \stabsf(\cF)$. This proves that the stabilizer subfield of $\cF$ is its best friend. Finally, by using the condition $1\in \cF_1$ together with Proposition \ref{prop: stab+ flag}, we obtain the inclusion
		$$
		\bbF_{q^m} \subseteq \stabsf(\cF) \subseteq  \stabsf(\cF_1) \subseteq \cF_1.
		$$ 
	\end{proof}
	\begin{remark}\label{rem: BF=interseccion BF_i y 1 in F_1}
		Observe that all flags in the code $\orb(\cF)$ have the same best friend. In particular, since $\orbbeta(\cF)\subseteq\orb(\cF)$, flags in a $\beta$-cyclic orbit flag code have all the same best friend for every $\beta\in\bbF_{q^n}^\ast$. Hence, we say that $\stabsf(\cF)$ is the best friend of every $\orbbeta(\cF)$.
	\end{remark}
	
	As stated in the proof of Proposition \ref{prop: stab+ flag}, (see equation (\ref{stab+ de un flag como intersección})), the stabilizer subfield of the cyclic flag code $\orb(\cF)$ can be computed as the intersection of the ones of its projected codes. Combining this with Proposition \ref{prop: stab+ es el best friend del flag}, we obtain the next result.
	\begin{corollary}\label{cor: BF flag is the intersection}
		Let $\cF=(\cF_1, \ldots, \cF_r)$ be a flag on $\bbF_{q^n}$. Then its best friend is the intersection of the ones of its subspaces. Moreover, if $1\in\cF_1$, every friend of the flag $\cF$ is contained in $\cF_1$.
	\end{corollary}
	It is clear that the best friend of a flag is a subfield of the ones of its subspaces. However, while the subspaces in a flag are nested, their respective best friends might not form a sequence of nested subfields as we can see in the following example.
	\begin{example}\label{ex:no nested best friends}
		Take $q$ a prime power and the flag of type $(2,3)$ on $\bbF_{q^4}$ given by $\cF=(\bbF_{q^2}, \bbF_{q^2}+\bbF_{q}\alpha)$, where $\alpha$ denotes a primitive element of $\bbF_{q^4}$. In this case, the best friend of $\cF_1$ is precisely $\bbF_{q^2}$ whereas, since $\gcd(3,4)=1$, the best friend of $\cF_2$ is the ground field $\bbF_q$.
	\end{example}
	
	As it happens in the subspace codes setting, knowing the best friend of a cyclic orbit flag code gives relevant information about the code parameters as we will see below.
	
	\section{Cyclic orbit flag codes with fixed best friend}\label{sec: prescribed BF}

	This section is devoted to the study of cyclic orbit flag codes on $\bbF_{q^n}$ generated by flags with the subfied $\bbF_{q^m}$ as their best friend. From now on, the integer $m$ will denote a divisor of $n$. Let us first see how the close relationship between the best friend of a flag and its stabilizer allows us to compute the size of the generated cyclic or $\beta$-cyclic orbit flag code. The next result follows  from (\ref{cardinality cyclic orbit flag code}) and Proposition \ref{prop: stab+ es el best friend del flag}.
	
	\begin{proposition}
		\label{prop: cardinality and best friend}
		Let $\cF=(\cF_1, \ldots, \cF_r)$ be a flag on $\bbF_{q^n}$ and $\beta \in \bbF_{q^n}^*$. Assume that $\bbF_{q^m}$ is the best friend of $\cF$. Then 
		$$
		|\orbbeta(\cF)|=\frac{|\beta|}{|\langle\beta\rangle \cap \bbF_{q^m}^\ast|}.
		$$
		In particular, if $\beta$ is a primitive element of $\bbF_{q^n}$, it holds $|\orb(\cF)|=\frac{q^n-1}{q^m-1}$.
	\end{proposition}

	\begin{remark}
		It is well known that any orbit coming from the action of a group can be partitioned into a set of orbits when we restrict the action to a subgroup. These orbits may have different cardinality in general. However, the cardinality of the code $\orbbeta(\cF)$ just depends on $|\beta|$ and   the best friend of $\cF$. Moreover, since all the flags in $\orb(\cF)$ have the same best friend, we have that $|\orbbeta(\cF')|=|\orbbeta(\cF)|$ for every $\cF'\in\orb(\cF)$. We conclude that, for any $\beta \in  \bbF_{q^n}^*$ the code $\orb(\cF)$ can be partitioned into a set of $\beta$-cyclic subcodes, all of them with the same cardinality. 
	\end{remark}
	
	Proposition \ref{prop: cardinality and best friend} leads to a characterization of $\beta$-cyclic orbit flag codes whose size coincides with the order or the acting group. 
	\begin{corollary}
		Let $\cF$ be a flag on $\bbF_{q^n}$ with $\bbF_{q^m}$ as its best friend and consider $\beta\in\bbF_{q^n}^\ast$. Then $|\orbbeta(\cF)|=|\beta|$ if, and only if $|\beta|$ and $q^m-1$ are coprime. In particular, this equality always holds if $q=2$ and $m=1$.
	\end{corollary}

		Having the subfield $\bbF_{q^m}$ as best friend yields a condition on the type vector of a flag, as well as a description of the structure of all the flags in its $\beta$-cyclic orbit flag code in terms of $\bbF_{q^m}$. Let $\cF=(\cF_1, \ldots, \cF_r)$ be a flag of type $(t_1, \ldots, t_r)$ on $\bbF_{q^n}$ with  $\bbF_{q^m}$ as its best friend. Hence, $\bbF_{q^m}$ must be a friend of all its subspaces and $m$ divides every dimension in the type vector. Consequently, we can write $t_i=m s_i$ for $i=1, \ldots, r$,  where $1\leq s_1 < s_2 < \dots < s_r < s = \frac{n}{m}$. On the other hand, the nested structure of the flag $\cF$ allows us to find linearly independent elements $a_1, \ldots, a_{s_r} \in \bbF_{q^{n}}$ (over $\bbF_{q^m}$) such that, for every $1\leq i\leq r$, we have
			$$
			\cF_i = \bigoplus_{j=1}^{s_i}  \bbF_{q^m} a_{j}.
			$$
		In particular, observe that if $m$ is a dimension in the type vector, then $s_1=1$ and the cyclic orbit code $\orb(\cF_1)$ is the $m$-spread of $\bbF_{q^n}$ described in (\ref{def: spread Anna-Lena}). Moreover, if $1\in\cF_1$, this subspace must be the subfield $\bbF_{q^m}$.

	Concerning the distance of $\beta$-cyclic orbit flag codes, as in the constant dimension codes framework, we can also deduce some estimates from the knowledge of the best friend.

	\begin{proposition}\label{prop: distance bounds BF}
		Let $\cF$ be a flag of type $(ms_1, \ldots, ms_r)$ on $\bbF_{q^n}$ with the subfield $\bbF_{q^m}$ as its best friend and take $\beta\in\bbF_{q^n}^\ast.$  Then $d_f(\orbbeta(\cF))=0$ if, and only if, $\beta\in\bbF_{q^m}^\ast$. Out of this case, $2m$ divides $d_f(\orbbeta(\cF))$ and it holds
		\begin{equation}\label{eq: distance bounds}
			2m \leq d_f(\orbbeta(\cF))  \leq  2m \left( \sum_{s_i \leq \lfloor \frac{s}{2}\rfloor} s_i + \sum_{s_i > \lfloor \frac{s}{2}\rfloor} (s-s_i) \right).  
		\end{equation}
	\end{proposition}
	\begin{proof}
Assume that $d_f(\orbbeta(\cF))=0$ or, equivalently, that $\orbbeta(\cF)=\{\cF\}$. This happens if, and only if, $\beta$ stabilizes the flag $\cF$, i.e., if $\beta\in\stab(\cF)=\bbF_{q^m}^\ast$.

Take now $\beta\in\bbF_{q^n}^\ast\setminus\bbF_{q^m}^\ast$. By the definition of best friend of the flag $\cF$, it follows that $\bbF_{q^m}$ is a friend of every subspace $\cF_i$.  This implies that, for every $1\leq i\leq r$, subspaces in $\orbbeta(\cF_i)$ are vector spaces over $\bbF_{q^m}$. Take a flag $\cF'$ in $\orbbeta(\cF)\setminus\{\cF\}.$ Since, for every $1\leq i \leq r,$ both $\cF_i$, $\cF'_i$, and hence  $\cF_i\cap\cF_i'$, are vector spaces over $\bbF_{q^{m}}$, the value $m$ divides their dimensions (over $\bbF_q$). Taking into account that $d_S(\cF_i, \cF_i') = 2(\dim(\cF_i)- \dim(\cF_i\cap\cF'_i))$, we conclude that $2m$ divides $d_S(\cF_i, \cF'_i)$ for every  $1\leq i\leq r$. Consequently, the value $2m$ also divides $d_f(\cF, \cF') = \sum_{i=1}^r d_S(\cF, \cF'),$ for every choice of $\cF'\in \orbbeta(\cF)\setminus\{\cF\}$. In particular, $2m$ divides $d_f(\orbbeta(\cF))$ and it is a lower bound for it. At the same time, if we consider the general upper bound for the distance of flag codes of type $(ms_1, \dots, ms_r)$ on $\bbF_{q^n}$ given in (\ref{eq: dist max flags}), taking into account that $n=ms$, we obtain the result.
	\end{proof}

	\begin{remark}
		Notice that for every $\beta\in\bbF_{q^n}^\ast$, it holds $\orbbeta(\cF) \subseteq \orb(\cF)$. Then it follows $d_f(\orbbeta(\cF))\geq d_f(\orb(\cF))$ except for $\beta\in\stab(\cF)=\bbF_{q^m}^\ast$. However, not every $\beta$ allows us to improve the distance with respect to the one of $\orb(\cF)$. We can  appreciate this fact in the next example.
	\end{remark}
	
	\begin{example}
		Take $q$ a prime power and $\alpha$ a primitive element of $\bbF_{q^{6}}$. Consider $\cF$ a flag of type $(1,4)$ on $\bbF_{q^{6}}$ with subspaces
		$$
		\cF_1=\bbF_{q} \ \text{and} \ \cF_2=\bbF_{q^{2}}+\bbF_{q^{2}}\alpha.
		$$
		Notice that, since $\gcd(1,4,6)=1$, by application of Lemma \ref{lem: BF divides dimensions}, $\bbF_q$ is the best friend of $\cF.$ Clearly, it is the best friend of $\cF_1$ as well. Concerning $\cF_2$, observe that $\bbF_{q^{2}}$ is one of its friends. Hence, its best friend is a subfield of $\bbF_{q^{6}}$ containing $\bbF_{q^{2}}$. We conclude that $\bbF_{q^2}$ is the best friend of $\cF_2$.
		The cyclic orbit flag code $\orb(\cF)$ contains exactly $\frac{q^6-1}{q-1}$ flags and we have $d_f(\orb(\cF))=2$. It suffices to see that, for every $\beta\in\bbF_{q^2}^\ast\setminus\bbF_q^\ast \subset \bbF_{q^6}^\ast$, it holds $\cF_2=\cF_2\beta$ and
		$$
		d_f(\cF, \cF\beta)= d_S(\cF_1, \cF_1\beta)=2.
		$$
		Observe that this is the minimum possible distance fixed the best friend $\bbF_q$. Now, if we consider the subgroup $\langle\gamma\rangle=\bbF_{q^2}^\ast$, the subcode  $\mathrm{Orb}_\gamma(\cF)$ has cardinality $\frac{q^2-1}{q-1}=q+1$ and the same argument above gives that $d_f(\mathrm{Orb}_\gamma(\cF))=2$. In this case, $\mathrm{Orb}_\gamma(\cF)$ does not have a better distance than $\orb(\cF)$. Take now $\delta\in\bbF_{q^6}^\ast$ a generator of $\bbF_{q^3}^\ast$, then the $\delta$-cyclic flag code generated by $\cF$ contains $\frac{q^3-1}{q-1}=q^2+q+1$ flags. To compute its distance, observe that 
		$$
		\mathrm{Stab}_\delta(\cF_2)= \langle\delta\rangle \cap \bbF_{q^2}^\ast =   \bbF_{q^3}^\ast \cap \bbF_{q^2}^\ast =\bbF_q^\ast = \mathrm{Stab}_\delta(\cF_1) = \mathrm{Stab}_\delta(\cF).
		$$
		Hence, for every $\delta^i\notin\mathrm{Stab}_\delta(\cF)$ it holds $\cF_j\neq\cF_j\delta^i$, for $j=1,2$. On the one hand, we have $d_S(\bbF_q, \bbF_q\delta^i)=2.$ On the other hand, as $\bbF_{q^2}$ is the best friend of $\cF_2$, the value $d_S(\cF_2, \cF_2\delta^i)$ is a multiple of $4$. Since the maximum possible distance between $4$-dimensional subspaces of $\bbF_{q^6}$ is precisely $2(6-4)=4$, it follows that $d_S(\cF_2, \cF_2\delta^i)=4$. As a result, $d_f(\cF, \cF\delta^i)=6$ for all $\delta^i\in\langle\delta\rangle\setminus \stab_\delta(\cF)$ and we conclude that $$d_f(\mathrm{Orb}_\delta(\cF))=6 >2=d_f(\orb(\cF)).$$
	\end{example}
	
	\begin{remark}
		Observe that the upper bound for the distance given in (\ref{eq: distance bounds}) coincides with the general bound for the flag distance given in (\ref{eq: dist max flags}). However, in Subsection \ref{subsec:optimum distance cyclic codes}, we will see that, in our scenario,  not every type vector is compatible with attaining this upper bound. On the other hand, the lower bound for the distance of a  $\beta$-cyclic flag code having $\bbF_{q^m}$ as its best friend obtained in (\ref{eq: distance bounds}) coincides with the one given in Proposition \ref{prop: stab+ es best friend} for  cyclic (subspace) codes having the same best friend. The previous example shows that this lower bound can also be attained by $\beta$-cyclic obit flag codes of length at least two. Let us see another situation where the generating flag has a special form.
	\end{remark}
	
	\begin{example}\label{ex: minimim distance flags}
		Let $\cF=(\bbF_{3^2}, \bbF_{3^4})$ be the flag of type $(2,4)$ on $\bbF_{3^8}$ defined in Example \ref{ex: inclusion estricta stabsfbeta} and consider the cyclic orbit flag code $\orb(\cF)$. Observe that, as stated in \ref{ex: inclusion estricta stabsfbeta}, the best friend of the flag $\cF$ is the subfield $\bbF_{3^2}$. Moreover, $\stab(\cF)=\stab(\cF_1)=\bbF_{3^2}$ and $\stab(\cF_2)=\bbF_{3^4}$.
		Now, if $\alpha$ denotes a primitive element of $\bbF_{3^8}$,  the power $\alpha^{82}$ is also a primitive element of the subfield $\bbF_{3^4}$. Hence, $\alpha^{82}$ clearly lies in $\bbF_{3^4}^\ast\setminus \bbF_{3^2}^\ast$. As a result, the flags $\cF$ and $\cF\alpha^{82}$ are different codewords in $\orb(\cF)$ whereas we have the subspaces equality $\cF_2=\cF_2\alpha^{82}$. It follows that
		$$
		d_f(\orb(\cF))\leq d_f(\cF, \cF\alpha^{82}) = d_S(\bbF_{3^2}, \bbF_{3^2}\alpha^{82})= 4,
		$$
		which is the minimum possible distance between subspaces of dimension one over $\bbF_{3^2}$. Hence, we conclude that $d_f(\orb(\cF))=4$.
	\end{example} 
	
	Notice that in the previous example the two subspaces of the generating flag are nested subfields of a given finite field. This example gives rise to the definition of a family of cyclic orbit flag codes inspired by the towers of subfields of $\bbF_{q^n}$.
	
	\subsection{Galois cyclic flag codes}\label{subsec:Galois codes}
	
	Let $1\leq t_1 < \dots < t_r < n$ be a sequence of divisors of $n$ such that $t_i$ divides $t_{i+1}$, for $1\leq i \leq r-1$. 
	
	\begin{definition}  We define the \emph{Galois flag} of type $(t_1, \dots, t_r)$ on $\bbF_{q^n}$ as the flag given by the sequence of nested subfields $(\bbF_{q^{t_1}}, \dots, \bbF_{q^{t_r}})$. For every $\beta\in \bbF_{q^n}^\ast$, the $\beta$-cyclic orbit flag code generated by this flag is called the \textit{Galois $\beta$-cyclic flag code} of type $(t_1, \dots, t_r)$. When $\beta$ is primitive, we just say \emph{Galois cyclic flag code.}
	\end{definition}
	
	\begin{remark}
		Notice that, for each subgroup $\langle\beta\rangle\subseteq\bbF_{q^n}^\ast$,  there is just one Galois $\beta$-cyclic flag code for each type vector satisfying the condition above. In contrast, the Galois $\beta$-cyclic flag code of a fixed type can be generated by different flags consisting of sequences of subspaces, not necessarily fields. Nevertheless, if we impose the condition $1\in\cF_1$, only the Galois flag of type $(t_1, \dots, t_r)$ can generate the Galois $\beta$-cyclic flag code of this type.
		
		Given the Galois flag $\cF$ of type vector $(t_1, \dots, t_r)$, it is clear that its $i$-th subspace is its own best friend. Hence, contrary to what happens for general flags (see Example \ref{ex:no nested best friends}), the best friends of the Galois flag subspaces form a sequence of nested subfieds. As a consequence, the first subfield $\bbF_{q^{t_1}}$ is the best friend of the Galois flag of type $(t_1, \dots, t_r)$ and, in order to construct Galois $\beta$-cyclic flag codes with the subfield $\bbF_{q^m}$ as its best friend, it suffices to consider a sequence of suitable divisors $(t_1, \dots, t_r)$ starting at $t_1=m$.
	\end{remark}
	
	Let us start focusing on Galois cyclic flag codes ($\beta$ primitive). According to Proposition \ref{prop: cardinality and best friend}, the cardinality of the Galois cyclic flag code of type $(t_1, \dots, t_r)$ is $c_1=(q^n-1)/(q^{t_1}-1)$ whereas its distance is $2t_1$. In particular, its $i$-projected code contains exactly $c_i=(q^n-1)/(q^{t_i}-1)$ subspaces and has subspace distance equal to $2t_i$. In spite of the fact that the distance of Galois cyclic flag codes is the smallest possible for cyclic orbit flag codes with a fixed best friend, the kaleidoscopic algebraic structure of nested spreads inside them is remarkable and deserves to be pointed out.
	
	\begin{theorem}\label{theo: Galois flag codes structure}
		Let $\cF=(\bbF_{q^{t_1}}, \dots, \bbF_{q^{t_r}})$ be the Galois flag of type $(t_1, \ldots, t_r)$ on the field $\bbF_{q^n}$ and $\orb(\cF)$ the associated Galois cyclic flag code. Consider $\alpha$ and $\alpha_i$ respective primitive elements of the fields $\bbF_{q^n}$ and $\bbF_{q^{t_i}}$, for $1\leq i\leq r$. Then it holds:
		\begin{enumerate}
			\item Each projected code of $\orb(\cF)$ is a $t_i$-spread of $\bbF_{q^n}$. 
			\item The $\alpha_j$-cyclic orbit code $\mathrm{Orb}_{\alpha_j}(\bbF_{q^{t_i}}\alpha^l)$ is a $t_i$-spread of the subspace $\bbF_{q^{t_j}}\alpha^l$, for every $i<j\leq r$ and $0\leq l \leq c_j-1$, where  $c_j=(q^n-1)/(q^{t_j}-1)$. 
		\end{enumerate}
		
	\end{theorem}
	
	\begin{proof}
		Observe that, by the definition of Galois cyclic flag code, the $i$-projected code $\orb(\cF_i)= \orb(\bbF_{q^{t_i}})$ is the $t_i$-spread of the field $\bbF_{q^n}$ described in (\ref{def: spread Anna-Lena}). The same argument allows us to state that, for every $i < j \leq r$, the $\alpha_j$-cyclic orbit code $\mathrm{Orb}_{\alpha_j}(\bbF_{q^{t_i}})$ is a $t_i$-spread of $\bbF_{q^{t_j}}$ as well. Moreover, since the subspace distance is invariant by the multiplicative action of $\bbF_{q^n}^\ast = \langle\alpha\rangle$ on subspaces, we have that $\mathrm{Orb}_{\alpha_j}(\bbF_{q^{t_i}}\alpha^l)$ is also a $t_i$-spread of the vector space $\bbF_{q^{t_j}}\alpha^l$, for every $0\leq l \leq q^n-2$. Now, taking into account that $\alpha^{c_j}$ is a primitive element of $\bbF_{q^{t_j}}$, we have that $\langle\alpha_j\rangle = \langle\alpha^{c_j}\rangle = \bbF_{q^{t_j}}^\ast$ and $\bbF_{q^{t_j}}=\bbF_{q^{t_j}}\alpha^{c_j}$. This fact allows us to restrict ourselves to exponents $0\leq l\leq c_j-1$.
	\end{proof}
	
	\begin{remark}
		Note that Theorem \ref{theo: Galois flag codes structure}, describes a striking cyclic spreads gear. First, every projected code of a Galois cyclic flag code is a spread. Then, every codeword in the $j$-projected code $\mathrm{Orb}(\bbF_{q^{t_j}}),$ i.e., every subspace of the form $\bbF_{q^{t_j}}\alpha^l$, is partitioned into the subspaces of the $\alpha_j$-cyclic orbit code $\mathrm{Orb}_{\alpha_j}(\bbF_{q^{t_i}}\alpha^l)$ if  $i<j\leq r$. Thereby, we have that $\mathrm{Orb}_{\alpha_j}(\bbF_{q^{t_i}}\alpha^l)$ is a $t_i$-spread of $\bbF_{q^{t_j}}\alpha^l$ for every value $0 \leq l\leq c_j-1$ and also a partial spread of dimension $t_i$ of the field $\bbF_{q^n}$. Finally, the union of all these orbits
		$$
		\dot\bigcup_{l=0}^{c_j-1} \mathrm{Orb}_{\alpha_j}(\bbF_{q^{t_i}}\alpha^l)
		$$
		gives us back the $t_i$-spread $\orb(\bbF_{q^{t_i}})=\orb(\cF_i)$. In other words, Galois cyclic flag codes provide collections of nested spreads that respect the orbital structure induced by the action of $\langle\alpha\rangle$ on flags.
	\end{remark}

	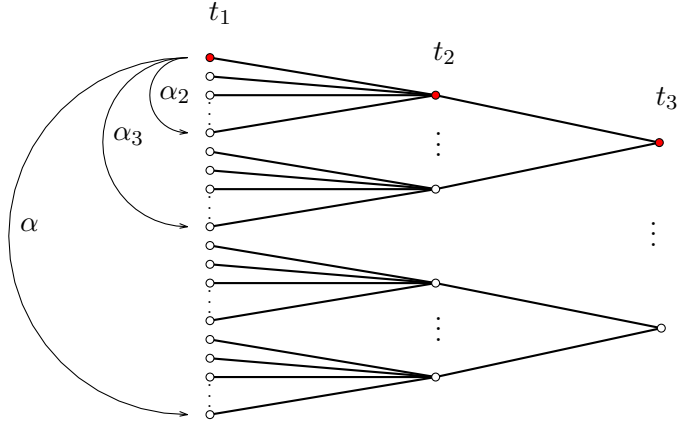
\begin{figure}[H]
		\begin{tikzpicture}[line cap=round,line join=round,>=triangle 45,x=1cm,y=1cm]
			\clip(-2,1) rectangle (9,8);
			\draw [line width=0.8pt] (2,6.75)-- (5,6.25);
			\draw [line width=0.8pt] (5,6.25)-- (2,6.5);
			\draw [line width=0.8pt] (2,6.25)-- (5,6.25);
			\draw [line width=0.8pt] (5,6.25)-- (2.00132,5.75);
			\draw [line width=0.8pt] (5,6.25)-- (7.973362657377393,5.619019447339418);
			\draw [line width=0.8pt] (7.973362657377393,5.619019447339418)-- (5,5);
			\draw [line width=0.8pt] (2,5.5)-- (5,5);
			\draw [line width=0.8pt] (5,5)-- (2,5.25);
			\draw [line width=0.8pt] (5,5)-- (2,5);
			\draw [line width=0.8pt] (2,2.5+1.75)-- (5,2+1.75);
			\draw [line width=0.8pt] (2,2+1.75)-- (5,2+1.75);
			\draw [line width=0.8pt] (2,1.5+1.75)-- (5,2+1.75);
			\draw [line width=0.8pt] (5,0.75+1.75)-- (2,1.25+1.75);
			\draw [line width=0.8pt] (5,0.75+1.75)-- (2,1+1.75);
			\draw [line width=0.8pt] (5,0.75+1.75)-- (2,0.75+1.75);
			\draw [line width=0.8pt] (5,0.75+1.75)-- (2,0.25+1.75);
			\draw [line width=0.8pt] (5,2+1.75)-- (8,1.4005718048604943+1.75);
			\draw [line width=0.8pt] (8,1.4005718048604943+1.75)-- (5,0.75+1.75);
			\draw (4.85,6.1) node[anchor=north west] {$\vdots$};
			\draw (4.85,3.65) node[anchor=north west] {$\vdots$};
			\draw (7.709433808911543,4.911482843761254) node[anchor=north west] {$\vdots$};
			\draw (1.825,5.25) node[anchor=north west] {{\tiny $\vdots$}};
			\draw (1.825,6.5) node[anchor=north west] {{\tiny $\vdots$}};
			\draw [line width=0.8pt] (2,4.5)-- (5,5);
			\draw [line width=0.8pt] (2,2.25+1.75)-- (5,2+1.75);
			\draw (1.825,2.25+1.75) node[anchor=north west] {{\tiny $\vdots$}};
			\draw (1.825,2.75) node[anchor=north west] {{\tiny $\vdots$}};
			\draw (1.1927775368697477,6.5) node[anchor=north west] {$\alpha_2$};
			\draw (0.5760592776859816,5.95) node[anchor=north west] {$\alpha_3$};
	        \draw (-0.6573772406815505,4.75) node[anchor=north west] {$\alpha$};
			\draw [shift={(1.7,6.25)},line width=0.4pt]  plot[domain=1.5707963267948966:4.71238898038469,variable=\t]({1*0.5*cos(\t r)+0*0.5*sin(\t r)},{0*0.5*cos(\t r)+1*0.5*sin(\t r)});
			\draw (1.8506103466657648,7.625043184169825) node[anchor=north west] {$t_1$};
			\draw (4.831415266053967,7.090554026210561) node[anchor=north west] {$t_2$};
			\draw (7.791662910136044,6.494393042332921) node[anchor=north west] {$t_3$};
			\draw [shift={(1.7,5.625)},line width=0.4pt]  plot[domain=1.5707963267948966:4.71238898038469,variable=\t]({1*1.125*cos(\t r)+0*1.125*sin(\t r)},{0*1.125*cos(\t r)+1*1.125*sin(\t r)});
			\draw [shift={(1.7,4.375)},line width=0.4pt]  plot[domain=1.5707963267948966:4.71238898038469,variable=\t]({1*2.375*cos(\t r)+0*2.375*sin(\t r)},{0*2.375*cos(\t r)+1*2.375*sin(\t r)});
			
			\draw [line width=0.4pt] (1.6,5.79)-- (1.7,5.75)-- (1.6,5.73);
			\draw [line width=0.4pt] (1.6,4.54)-- (1.7,4.5)-- (1.6,4.48);
			\draw [line width=0.4pt] (1.6,2.04)-- (1.7,2)-- (1.6,1.98);
			
    		\begin{scriptsize} 
				\draw [fill=white] (5,2+1.75) circle (1.5pt);
				\draw [fill=red] (2,6.75) circle (1.5pt);
				\draw [fill=white] (2,6.25) circle (1.5pt);
				\draw [fill=white] (2,5.75) circle (1.5pt);
				\draw [fill=red] (5,6.25) circle (1.5pt);
				\draw [fill=white] (2,6.5) circle (1.5pt);
				\draw [fill=red] (7.973362657377393,5.619019447339418) circle (1.5pt);
				\draw [fill=white] (5,5) circle (1.5pt);
				\draw [fill=white] (2,5) circle (1.5pt);
				\draw [fill=white] (2,5.5) circle (1.5pt);
				\draw [fill=white] (2,5.25) circle (1.5pt);
				\draw [fill=white] (2,2.5+1.75) circle (1.5pt);
				\draw [fill=white] (2,2+1.75) circle (1.5pt);
				\draw [fill=white] (2,1.5+1.75) circle (1.5pt);
				\draw [fill=white] (2,1.25+1.75) circle (1.5pt);
				\draw [fill=white] (2,1+1.75) circle (1.5pt);
				\draw [fill=white] (2,0.75+1.75) circle (1.5pt);
				\draw [fill=white] (2,0.25+1.75) circle (1.5pt);
				\draw [fill=white] (5,0.75+1.75) circle (1.5pt);
				\draw [fill=white] (8,1.4005718048604943+1.75) circle (1.5pt);
				\draw [fill=white] (2,4.5) circle (1.5pt);
				\draw [fill=white] (2,2.25+1.75) circle (1.5pt);
			\end{scriptsize}
		\end{tikzpicture}
		\caption{Nested spread structure of a Galois cyclic flag code}\label{fig1}
	\end{figure}
	\noindent The previous figure represents the structure of the Galois cyclic flag code of a given type $(t_1, t_2, t_3)$. Vertices are subspaces, (directed) edges denote inclusions (from left to right) and flags are given by directed paths in the graph. Each column in the graph is a projected code and, by Theorem \ref{theo: Galois flag codes structure}, all of them are spreads of $\bbF_q^n$ of the corresponding dimensions. In addition, every subspace in the graph is partitioned into the set of its left adjacent vertices. On the other hand, the Galois flag $\cF=(\bbF_{q^{t_1}}, \bbF_{q^{t_2}}, \bbF_{q^{t_3}})$ is represented by the sequence of red vertices. Since $\stab(\cF)=\bbF_{q^{t_1}}^\ast= \langle \alpha_1\rangle$, the code $\mathrm{Orb}_{\alpha_1}(\cF)$ consists of the single element $\cF.$ In contrast, for $i=2, 3$, the code $\mathrm{Orb}_{\alpha_i}(\cF)$ is given by the set of flags in the graph marked by the round arrow labeled with $\alpha_i$. 
	
	\vspace{0.25cm}
	
	Take now an element $\beta\in \bbF_{q^n}^\ast$. Let $\cF$  be the Galois flag of type $(t_1, \dots, t_r)$ on $\bbF_{q^n}$ and consider the Galois $\beta$-cyclic flag code $\mathrm{Orb}_{\beta}(\cF)$. Since $\bbF_{q^{t_1}}$ is the best friend of $\cF$, it follows that $\stabbeta(\cF)=\langle\beta\rangle\cap\bbF_{q^{t_1}}^\ast$. Moreover, for every value of $1\leq i\leq r$, it holds $\stabbeta(\cF_i)=\langle\beta\rangle\cap\bbF_{q^{t_i}}^\ast.$ 
	As a result, we have the following sequence of nested subgroups of $\langle\beta\rangle$ 
	\begin{equation}\label{eq: Galois beta cyclic stab}
		\stabbeta(\cF) = \stabbeta(\cF_1) \subseteq \stabbeta(\cF_2) \subseteq \dots \subseteq \stabbeta(\cF_r) \subseteq \langle\beta\rangle.
	\end{equation}
	By means of Proposition \ref{prop: cardinality and best friend}, the cardinality of $\orbbeta(\cF)$ and the one of its $i$-projected code, for every $1\leq i\leq r$, are respectively
	$$
	|\orbbeta(\cF)|= \frac{|\beta|}{|\langle\beta\rangle\cap\bbF_{q^{t_1}}^\ast|} \ \text{and} \ |\orbbeta(\cF_i)|= \frac{|\beta|}{|\langle\beta\rangle\cap\bbF_{q^{t_i}}^\ast|}
	.
	$$
	Furthermore, from Theorem \ref{theo: Galois flag codes structure}, and taking into account that $\orbbeta(\cF)\subseteq\orb(\cF)$, we can derive the following result for the projected codes of a Galois $\beta$-cyclic flag code.
	
	\begin{corollary}\label{theo: Galois beta cyclic flag codes structure}
		Let $\cF=(\bbF_{q^{t_1}}, \dots, \bbF_{q^{t_r}})$ be the Galois flag of type $(t_1, \ldots, t_r)$ on the field $\bbF_{q^n}$ and take a nonzero element $\beta\in \bbF_{q^n}$. For each $1\leq i\leq r,$ we write $\beta_i$ to denote a generator of the cyclic subgroup $\langle\beta_i\rangle=\stabbeta(\bbF_{q^{t_i}})=\langle\beta\rangle\cap\bbF_{q^{t_i}}^\ast.$ Then the following statements hold:
		\begin{enumerate}
			\item The projected code $\orbbeta(\cF_i)$ is a partial spread of dimension $t_i$ of $\bbF_{q^n}$. 
			\item The $\beta_j$-cyclic orbit code $\mathrm{Orb}_{\beta_j}(\bbF_{q^{t_i}}\beta^l)$ is a partial spread of dimension $t_i$ of the subspace $\bbF_{q^{t_j}}\beta^l$, for every $i<j\leq r$ and $0\leq l \leq |\beta_j|-1$.
		\end{enumerate}
	\end{corollary}

	Concerning the distance of Galois $\beta$-cyclic flag codes, since they are subcodes of the Galois cyclic flag code of the same type, their distance might be better than $2t_1,$ apart from the case of the trivial subcode consisting just of the Galois flag, which has distance equal to zero. Actually, it is possible to determine the exact distance of a Galois $\beta$-cyclic flag code by checking the relationship between the subgroup $\langle \beta \rangle$ and the subfields $\bbF_{q^{t_i}}$ and vice versa, that is, if we choose a permitted distance, we can find a suitable subgroup (possibly not unique) to build a $\beta$-cyclic orbit Galois attaining such a distance. We state the precise conditions in the following result:

\begin{theorem}\label{theo: distance Galois beta cyclic}
		Let $\cF$ be the Galois flag of type $(t_1, \dots, t_r)$ and consider an element $\beta\in \bbF_{q^n}^\ast$. Then $d_f(\orbbeta(\cF)) \in \{ 0, 2t_1, 2(t_1+t_2),\dots, 2(t_1+t_2+\dots+t_r)\}$. Moreover,
		\begin{enumerate}
			\item $d_f(\orbbeta(\cF))= 0$ if, and only if, $\stabbeta(\cF_1)=\stabbeta(\cF_r)=\langle\beta\rangle$. \label{theo: distance Galois beta cyclic item 1}
			\item $d_f(\orbbeta(\cF))= 2\sum_{i=1}^r t_i$ if, and only if, $\stabbeta(\cF_1)=\stabbeta(\cF_r)\neq\langle\beta\rangle$. \label{theo: distance Galois beta cyclic item 2}
			\item $d_f(\orbbeta(\cF))= 2\sum_{i=1}^{j-1} t_i$  if, and only if, $\stabbeta(\cF_1)\neq\stabbeta(\cF_r)$ and $j\in\{2, \dots, r\}$ is the minimum index such that $\stabbeta(\cF_1) \subsetneq \stabbeta(\cF_j).$ \label{theo: distance Galois beta cyclic item 3}
		\end{enumerate}
\end{theorem}
\begin{proof}

Recall that for every choice of $\beta$, the projected codes of $\orbbeta(\cF)$ are partial spreads. As a result, for every $0\leq l\leq |\beta|-1$, we have that $d_S(\cF_j, \cF_j\beta^l)\in\{0, 2t_j\}$. Moreover, $d_S(\cF_j, \cF_j\beta^l)=0$ holds if, and only if, $\beta^l\in\stabbeta(\cF_j)$. In this case, since $\stabbeta(\cF_j)\subseteq \dots \subseteq \stabbeta(\cF_r)$ by (\ref{eq: Galois beta cyclic stab}), we have $d_S(\cF_i, \cF_i\beta^l)=0,$ for every $j\leq i\leq r$. Hence, distances between flags in $\orbbeta(\cF)$ belong to the set $\{0, 2t_1, 2(t_1+t_2),\dots, 2(t_1+t_2+\dots+t_r)\}.$ Let us see that all of them can be reached, by showing (\ref{theo: distance Galois beta cyclic item 1}), (\ref{theo: distance Galois beta cyclic item 2}) and (\ref{theo: distance Galois beta cyclic item 3}).

\begin{enumerate}
\item	As proved in Proposition \ref{prop: distance bounds BF}, we have $d_f(\orbbeta(\cF))=0$ if, and only if, $\beta\in\bbF_{q^{t_1}}^\ast=\stab(\cF)$ or, by using  (\ref{eq: estabilizador flag}), $\beta\in\stab(\cF_i)$ for all $1\leq i\leq r$. Since $\stabbeta(\cF_i)=\langle\beta\rangle\cap\stab(\cF_i)$ is always a subgroup of $\langle\beta\rangle$, the previous condition is equivalent to $\stabbeta(\cF_i)=\langle\beta\rangle$, for every $1\leq i\leq r$. Hence, by (\ref{eq: Galois beta cyclic stab}), we just need to check the equality $\stabbeta(\cF_1)=\stabbeta(\cF_r)=\langle\beta\rangle.$
\end{enumerate}
In the remaining cases, $\stabbeta(\cF)$ must be a proper subgroup of $\langle\beta\rangle$.
\begin{enumerate}
\item[(2)] Assume now that $d_f(\orbbeta(\cF))= 2\sum_{i=1}^r t_i$. Hence, for every $\beta^l\in\langle\beta\rangle\setminus\stabbeta(\cF)$, it must hold $d_S(\cF_i,\cF_i\beta^l)= 2t_i$, for all $1\leq i\leq r$. This happens if, and only if, $\beta^l\notin\stabbeta(\cF_i)$ for every $1\leq i\leq r$. As a consequence, $\stabbeta(\cF_i)\subseteq\stabbeta(\cF)$. On the other hand, by (\ref{eq: estabilizador flag}), we conclude that $\stabbeta(\cF)=\stabbeta(\cF_i)$ for every $1\leq i\leq r$. Again, since these stabilizer subgroups are nested, this condition is equivalent to $\stabbeta(\cF_1)=\stabbeta(\cF_r).$
		
\item[(3)] Consider the case $d_f(\orbbeta(\cF))= 2\sum_{i=1}^{j-1} t_i$ for some $2\leq j\leq r$. In other words, there exists some $\beta^l\in\langle\beta\rangle\setminus\stabbeta(\cF)$ such that
$$
d_f(\orbbeta(\cF))= d_f(\cF, \cF\beta^l)=2\sum_{i=1}^{j-1} t_i.
$$
This happens if, and only if
$$
d_S(\cF_i, \cF_i\beta^l)=\left\lbrace
\begin{array}{ccl}
2t_i &\text{if} & 1\leq i\leq j-1,\\
0    &\text{if} & j\leq i\leq r,
\end{array}
\right.
$$
or equivalently, if $\beta^l\in\langle\beta\rangle\setminus\stabbeta(\cF_i)$ for $1\leq i\leq j-1,$ and $\beta^l\in\stabbeta(\cF_i)$ for $j\leq i\leq r$. Hence, we conclude $$\stabbeta(\cF)=\stabbeta(\cF_1)=\dots=\stabbeta(\cF_{j-1})\subsetneq\stabbeta(\cF_j).$$
\end{enumerate}
\end{proof}

Graphically, Galois $\beta$-cyclic flag codes can be represented as subgraphs of the graph in Figure \ref{fig1}. In the next picture, flags in a Galois $\beta$-cyclic flag code are marked with black lines. In contrast, directed paths containing dotted edges represent flags in $\orb(\cF)\setminus\orbbeta(\cF)$. The index $j$ in Theorem \ref{theo: distance Galois beta cyclic} states that no flags in the code share subspaces of dimensions $t_i$, for $1\leq i\leq j-1,$ whereas there exist different flags having the same $j$-th subspace. At left, and example of Galois $\beta$-cyclic flag code with distance $2t_1$ ($j=2$). At right, the corresponding index and distance are $j=3$ and $2(t_1+t_2),$ respectively.

\begin{figure}[H]\label{fig2}
\begin{tikzpicture}[line cap=round,line join=round,>=triangle 45,x=0.75cm,y=1cm]
\clip(-2,1.75) rectangle (15,7);
\draw [line width=0.8pt] (2-2,6.75)-- (5-2,6.25);
\draw [line width=0.8pt] (5-2,6.25)-- (2-2,6.5);
\draw [line width=0.8pt, dotted] (2-2,6.25)-- (5-2,6.25);
\draw [line width=0.8pt, dotted] (5-2,6.25)-- (2-2,5.75);
\draw [line width=0.8pt] (5-2,6.25)-- (8-2,5.619019447339418);
\draw [line width=0.8pt] (8-2,5.619019447339418)-- (5-2,5);
\draw [line width=0.8pt] (2-2,5.5)-- (5-2,5);
\draw [line width=0.8pt] (5-2,5)-- (2-2,5.25);
\draw [line width=0.8pt, dotted] (5-2,5)-- (2-2,5);
\draw [line width=0.8pt] (2-2,2.5+1.75)-- (5-2,2+1.75);
\draw [line width=0.8pt, dotted] (2-2,2+1.75)-- (5-2,2+1.75);
\draw [line width=0.8pt, dotted] (2-2,1.5+1.75)-- (5-2,2+1.75);
\draw [line width=0.8pt] (5-2,0.75+1.75)-- (2-2,1.25+1.75);
\draw [line width=0.8pt] (5-2,0.75+1.75)-- (2-2,1+1.75);
\draw [line width=0.8pt, dotted] (5-2,0.75+1.75)-- (2-2,0.75+1.75);
\draw [line width=0.8pt, dotted] (5-2,0.75+1.75)-- (2-2,0.25+1.75);
\draw [line width=0.8pt] (5-2,2+1.75)-- (8-2,1.4005718048604943+1.75);
\draw [line width=0.8pt] (8-2,1.4005718048604943+1.75)-- (5-2,0.75+1.75);
\draw (4.85-2,6.1) node[anchor=north west] {$\vdots$};
\draw (4.85-2,3.65) node[anchor=north west] {$\vdots$};
\draw (7.709433808911543-2,4.911482843761254) node[anchor=north west] {$\vdots$};
\draw (1.825-2,5.25) node[anchor=north west] {{\tiny $\vdots$}};
\draw (1.825-2,6.5) node[anchor=north west] {{\tiny $\vdots$}};
\draw [line width=0.8pt, dotted] (2-2,4.5)-- (5-2,5);
\draw [line width=0.8pt] (2-2,2.25+1.75)-- (5-2,2+1.75);
\draw (1.825-2,2.25+1.75) node[anchor=north west] {{\tiny $\vdots$}};
\draw (1.825-2,2.75) node[anchor=north west] {{\tiny $\vdots$}};
\draw (1.8506103466657648-2,7.625043184169825) node[anchor=north west] {$t_1$};
\draw (4.831415266053967-2,7.090554026210561) node[anchor=north west] {$t_2$};
\draw (7.791662910136044-2,6.494393042332921) node[anchor=north west] {$t_3$};

\begin{scriptsize} 
\draw [fill=white] (5-2,2+1.75) circle (1.5pt);
\draw [fill=red] (2-2,6.75) circle (1.5pt);
\draw [fill=white] (2-2,6.25) circle (1.5pt);
\draw [fill=white] (2-2,5.75) circle (1.5pt);
\draw [fill=red] (5-2,6.25) circle (1.5pt);
\draw [fill=white] (2-2,6.5) circle (1.5pt);
\draw [fill=red] (8-2,5.619019447339418) circle (1.5pt);
\draw [fill=white] (5-2,5) circle (1.5pt);
\draw [fill=white] (2-2,5) circle (1.5pt);
\draw [fill=white] (2-2,5.5) circle (1.5pt);
\draw [fill=white] (2-2,5.25) circle (1.5pt);
\draw [fill=white] (2-2,2.5+1.75) circle (1.5pt);
\draw [fill=white] (2-2,2+1.75) circle (1.5pt);
\draw [fill=white] (2-2,1.5+1.75) circle (1.5pt);
\draw [fill=white] (2-2,1.25+1.75) circle (1.5pt);
\draw [fill=white] (2-2,1+1.75) circle (1.5pt);
\draw [fill=white] (2-2,0.75+1.75) circle (1.5pt);
\draw [fill=white] (2-2,0.25+1.75) circle (1.5pt);
\draw [fill=white] (5-2,0.75+1.75) circle (1.5pt);
\draw [fill=white] (8-2,1.4005718048604943+1.75) circle (1.5pt);
\draw [fill=white] (2-2,4.5) circle (1.5pt);
\draw [fill=white] (2-2,2.25+1.75) circle (1.5pt);
\end{scriptsize}

\draw [line width=0.8pt] (2+6,6.75)-- (5+6,6.25);
\draw [line width=0.8pt, dotted] (5+6,6.25)-- (2+6,6.5);
\draw [line width=0.8pt, dotted] (2+6,6.25)-- (5+6,6.25);
\draw [line width=0.8pt, dotted] (5+6,6.25)-- (2+6,5.75);
\draw [line width=0.8pt] (5+6,6.25)-- (8+6,5.619019447339418);
\draw [line width=0.8pt] (8+6,5.619019447339418)-- (5+6,5);
\draw [line width=0.8pt] (2+6,5.5)-- (5+6,5);
\draw [line width=0.8pt, dotted] (5+6,5)-- (2+6,5.25);
\draw [line width=0.8pt, dotted] (5+6,5)-- (2+6,5);
\draw [line width=0.8pt] (2+6,2.5+1.75)-- (5+6,2+1.75);
\draw [line width=0.8pt, dotted] (2+6,2+1.75)-- (5+6,2+1.75);
\draw [line width=0.8pt, dotted] (2+6,1.5+1.75)-- (5+6,2+1.75);
\draw [line width=0.8pt] (5+6,0.75+1.75)-- (2+6,1.25+1.75);
\draw [line width=0.8pt, dotted] (5+6,0.75+1.75)-- (2+6,1+1.75);
\draw [line width=0.8pt, dotted] (5+6,0.75+1.75)-- (2+6,0.75+1.75);
\draw [line width=0.8pt, dotted] (5+6,0.75+1.75)-- (2+6,0.25+1.75);
\draw [line width=0.8pt] (5+6,2+1.75)-- (8+6,1.4005718048604943+1.75);
\draw [line width=0.8pt] (8+6,1.4005718048604943+1.75)-- (5+6,0.75+1.75);
\draw (4.85+6,6.1) node[anchor=north west] {$\vdots$};
\draw (4.85+6,3.65) node[anchor=north west] {$\vdots$};
\draw (7.709433808911543+6,4.911482843761254) node[anchor=north west] {$\vdots$};
\draw (1.825+6,5.25) node[anchor=north west] {{\tiny $\vdots$}};
\draw (1.825+6,6.5) node[anchor=north west] {{\tiny $\vdots$}};
\draw [line width=0.8pt, dotted] (2+6,4.5)-- (5+6,5);
\draw [line width=0.8pt, dotted] (2+6,2.25+1.75)-- (5+6,2+1.75);
\draw (1.825+6,2.25+1.75) node[anchor=north west] {{\tiny $\vdots$}};
\draw (1.825+6,2.75) node[anchor=north west] {{\tiny $\vdots$}};
\draw (1.8506103466657648+6,7.625043184169825) node[anchor=north west] {$t_1$};
\draw (4.831415266053967+6,7.090554026210561) node[anchor=north west] {$t_2$};
\draw (7.791662910136044+6,6.494393042332921) node[anchor=north west] {$t_3$};

\begin{scriptsize} 
\draw [fill=white] (5+6,2+1.75) circle (1.5pt);
\draw [fill=red] (2+6,6.75) circle (1.5pt);
\draw [fill=white] (2+6,6.25) circle (1.5pt);
\draw [fill=white] (2+6,5.75) circle (1.5pt);
\draw [fill=red] (5+6,6.25) circle (1.5pt);
\draw [fill=white] (2+6,6.5) circle (1.5pt);
\draw [fill=red] (8+6,5.619019447339418) circle (1.5pt);
\draw [fill=white] (5+6,5) circle (1.5pt);
\draw [fill=white] (2+6,5) circle (1.5pt);
\draw [fill=white] (2+6,5.5) circle (1.5pt);
\draw [fill=white] (2+6,5.25) circle (1.5pt);
\draw [fill=white] (2+6,2.5+1.75) circle (1.5pt);
\draw [fill=white] (2+6,2+1.75) circle (1.5pt);
\draw [fill=white] (2+6,1.5+1.75) circle (1.5pt);
\draw [fill=white] (2+6,1.25+1.75) circle (1.5pt);
\draw [fill=white] (2+6,1+1.75) circle (1.5pt);
\draw [fill=white] (2+6,0.75+1.75) circle (1.5pt);
\draw [fill=white] (2+6,0.25+1.75) circle (1.5pt);
\draw [fill=white] (5+6,0.75+1.75) circle (1.5pt);
\draw [fill=white] (8+6,1.4005718048604943+1.75) circle (1.5pt);
\draw [fill=white] (2+6,4.5) circle (1.5pt);
\draw [fill=white] (2+6,2.25+1.75) circle (1.5pt);
\end{scriptsize}
\end{tikzpicture}
\caption{Two different Galois $\beta$-cyclic of type $(t_1, t_2, t_3).$}
\end{figure}
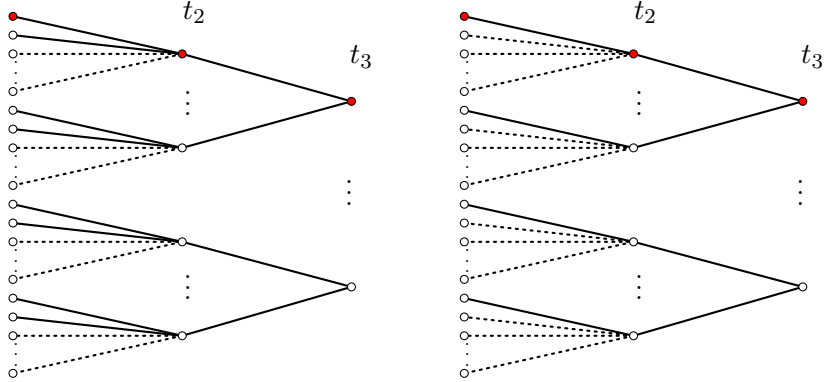
	
	Observe that Theorem \ref{theo: distance Galois beta cyclic} allows us to provide specific constructions of Galois $\beta$-cyclic flag codes with a prescribed distance just by choosing a suitable element $\beta\in\bbF_{q^n}^\ast$. Moreover, since  $\bbF_{q^n}^\ast=\langle\alpha\rangle$, being $\alpha$ a primitive element of $\bbF_{q^n}$, we can translate the above conditions on the stabilizers (w.r.t. $\beta$) in terms of suitable powers of $\alpha$ as follows. Given $\beta\in\bbF_{q^n}^\ast$, we can write $|\beta|=(q^n-1)/l$ for some divisor $l$ of $q^n-1$. Hence, by the uniqueness of subgroups of a given order of the cyclic group $\bbF_{q^n}^\ast$, it is clear that $\langle\beta\rangle=\langle\alpha^l\rangle$. In particular, if $c_i=(q^n-1)/(q^{t_i}-1)$, we have that $\bbF_{q^{t_i}}^\ast=\langle\alpha^{c_i}\rangle$, for every $1\leq i\leq r$. As a consequence, it holds $\stabbeta(\bbF_{q^{t_i}})=\langle\beta\rangle\cap\bbF_{q^{t_i}}^\ast= \langle\alpha^l\rangle\cap \langle\alpha^{c_i}\rangle  = \langle\alpha^{l_i}\rangle,$ where $l_i=\lcm(l, c_i)$. Moreover, given that each $c_{i+1}$ divides $c_i$, then $l_{i+1}$ divides $l_i$, for every $1\leq i\leq r-1$, and the sequence of nested stabilizers given in (\ref{eq: Galois beta cyclic stab}) becomes 
$$
\langle\alpha^{l_1}\rangle\subseteq \langle\alpha^{l_2} \rangle \subseteq \dots \subseteq \langle\alpha^{l_r}\rangle \subseteq \langle\alpha^l\rangle.
$$	
Now, since $l, c_1, \dots, c_r$ divide $q^n-1$, every exponent $l_i$ divides $q^n-1$ as well. Hence, the order of each stabilizer is $|\stabbeta(\bbF_{q^{t_i}})|=|\alpha^{l_i}|= \frac{q^n-1}{l_i}$, for every $1\leq i\leq r$. We can reformulate Theorem \ref{theo: distance Galois beta cyclic} as follows:

\begin{theorem}
Let $\cF$ be the Galois flag of type $(t_1, \dots, t_r)$ and consider $\beta\in \bbF_{q^n}^\ast$ such that $\langle\beta\rangle=\langle\alpha^l\rangle$ for some divisor $l$ of $q^n-1$. It holds:
		\begin{enumerate}
			\item $d_f(\orbbeta(\cF))= 0$ if, and only if, $l_1=l_r=l$.
			\item $d_f(\orbbeta(\cF))= 2\sum_{i=1}^r t_i$ if, and only if, $l_1=l_r\neq l$.
			\item $d_f(\orbbeta(\cF))= 2\sum_{i=1}^{j-1} t_i$ if, and only if, $l_1\neq l_r$ and $2\leq j\leq r$ is the minimum index such that $l_1\neq l_j$
		\end{enumerate}
\end{theorem}

\begin{example}
Take $\cF$ the Galois flag of type $(2,4,8)$ on $\bbF_{2^{16}}$ and let $\alpha$ be a primitive element of $\bbF_{2^{16}}.$ The following table shows the parameters of all possible Galois $\beta$-cyclic flag codes of this type. The sizes of the stabilizer subgroups (w.r.t. $\beta$) of the fields $\bbF_{2^2}$, $\bbF_{2^4}$ and $\bbF_{2^8}$ are given, together with the cardinality  and distance (just denoted by $d_\beta$) of $\orbbeta(\cF).$ 

\begin{table}[H]
\centering
\begin{small}
\begin{tabular}{ccccccc}
\hline
$\beta$       & $|\beta|$ & $|\stabbeta(\bbF_{2^{2}})|$ & $|\stabbeta(\bbF_{2^{4}})|$ & $|\stabbeta(\bbF_{2^{8}})|$ & $|\orbbeta(\cF)|$ & $d_\beta$      \\  \hline
$\alpha$           &  65535     & 3 & 15 & 255 &  21845 &  4  \\
$\alpha^3$         &  21845     & 1 &  5 &  85 &  21845 &  4  \\
$\alpha^5$         &  13107     & 3 &  3 &  51 &   4369 & 12  \\
$\alpha^{15}$      &   4369     & 1 &  1 &  17 &   4369 & 12  \\
$\alpha^{17}$      &   3855     & 3 & 15 &  15 &   1285 &  4  \\
$\alpha^{51}$      &   1285     & 1 &  5 &   5 &   1285 &  4  \\
$\alpha^{85}$      &    771     & 3 &  3 &   3 &    257 & 28  \\
$\alpha^{255}$     &    257     & 1 &  1 &   1 &    257 & 28  \\
$\alpha^{257}$     &    255     & 3 & 15 & 255 &     85 &  4  \\
$\alpha^{771}$     &     85     & 1 &  5 &  85 &     85 &  4  \\
$\alpha^{1285}$    &     51     & 3 &  3 &  51 &     17 & 12  \\
$\alpha^{3855}$    &     17     & 1 &  1 &  17 &     17 & 12  \\
$\alpha^{4369}$    &     15     & 3 & 15 &  15 &      5 &  4  \\
$\alpha^{13107}$   &      5     & 1 &  5 &   5 &      5 &  4  \\
$\alpha^{21845}$   &      3     & 3 &  3 &   3 &      1 &  0  \\
1                  &      1     & 1 &  1 &   1 &      1 &  0  \\ \hline
\end{tabular}
\end{small}
\caption{Parameters of all Galois $\beta$-cyclic flag codes of type $(2,4,8)$ on $\bbF_{2^{16}}$.} 
\end{table}

Clearly, different subgroups of $\bbF_{q^n}^\ast$ can provide the same code. For instance, the subgroup $\langle\alpha^3\rangle$ gives the Galois cyclic flag code $\orb(\cF)$. We have also $\orb_{\alpha^{5}}(\cF)=\orb_{\alpha^{15}}(\cF)$ or $\orb_{\alpha^{85}}(\cF)=\orb_{\alpha^{255}}(\cF)$, among other possibilities.

\end{example}

	\begin{remark}
As proved in the previous theorem, the Galois $\beta$-cyclic code of type $(t_1, \dots, t_r)$ attains the maximum possible distance for its type if, and only if, it holds
		\begin{equation}\label{eq: condition Galois ODFC}
		\stabbeta(\cF_1)= \stabbeta(\cF_r)  \subsetneq \langle\beta\rangle .
		\end{equation} 
In other words, if condition (\ref{eq: condition Galois ODFC}) is satisfied, we can build an optimum distance flag code with $\bbF_{q^{t_1}}$ as its best friend. This fact drives us to investigate cyclic orbit flag codes with the maximum possible distance and fixed best friend when the generating flag is not necessarily a Galois flag.
\end{remark}

	\subsection{Optimum distance cyclic orbit flag codes}\label{subsec:optimum distance cyclic codes}
	
	This subsection is devoted to the study of flag codes on $\bbF_{q^n}$ reaching the maximum distance and being also $\beta$-cyclic orbit flag codes with a prescribed best friend $\bbF_{q^m}$.  To tackle this problem, we have to take into account first that, in particular, optimum distance flag codes must be disjoint as proved in \cite{CasoPlanar}. 
	Recall that a flag code $\cC$ of type $(t_1, \ldots, t_r)$ is said to be \emph{disjoint} if $|\cC|=|\cC_1|= \dots= |\cC_r|.$ In our specific context, we have that a $\beta$-cyclic flag code $\orbbeta(\cF)$ is disjoint if, and only if, 
	$$
	\frac{|\beta|}{|\stabbeta(\cF)|}= \frac{|\beta|}{|\stabbeta(\cF_1)|}= \cdots= \frac{|\beta|}{|\stabbeta(\cF_r)|}
	$$
	or, equivalently, if all the stabilizers $\stabbeta(\cF), \stabbeta(\cF_1), \ldots, \stabbeta(\cF_r)$ have the same order. In fact, by the uniqueness of subgroups of a cyclic group, all these stabilizers must coincide. Moreover, by using (\ref{eq: estabilizador flag}), we have the next result:

	\begin{proposition}\label{prop: disjunto iff estabilizadores coinciden}
		The following statements are equivalent:
		\begin{enumerate}
			\item $\orbbeta(\cF)$ is a disjoint flag code, \label{condición 1}
			\item $\stabbeta(\cF)=\stabbeta(\cF_1)=\dots= \stabbeta(\cF_r)$.\label{condición 2}
			\item $\stabbeta(\cF_1)=\dots= \stabbeta(\cF_r)$.\label{condición 3}
		\end{enumerate}
	\end{proposition}
	In light of Propositions \ref{prop: stab+ flag} and \ref{prop: stab+ es el best friend del flag}, the best friend of a flag $\cF$ can be computed as $\stabsf(\cF)=\stab(\cF)\cup\{0\}$. Similarly, the best friend of its subspaces are given by $\stabsf(\cF_i)=\stab(\cF_i)\cup\{0\}$. The next result leads directly a characterization of disjoint $\beta$-cyclic orbit flag codes in terms of $\beta$ and the best friends of the generating flag and its subspaces.
\begin{proposition}\label{prop: disjunto best friend}
Let $\cF=(\cF_1, \ldots, \cF_r)$ be a flag on $\bbF_{q^n}$ with $\bbF_{q^m}$ as its best friend and take  $\beta\in\bbF_{q^n}^\ast$.  If $\bbF_{q^{m_i}}$ denotes the best friend of $\cF_i$, then the $\beta$-cyclic orbit code $\orbbeta(\cF)$ is disjoint if, and only if
		$$
		\langle\beta\rangle\cap\bbF_{q^m}^\ast= \langle\beta\rangle\cap\bbF_{q^{m_1}}^\ast= \dots = \langle\beta\rangle\cap\bbF_{q^{m_r}}^\ast.
		$$
In particular, the cyclic orbit flag code $\orb(\cF)$ is disjoint if, and only if, all the subspaces in the flag have the field $\bbF_{q^m}$ as their best friend. 
	\end{proposition}
	\begin{proof}
	By means of Proposition \ref{prop: disjunto iff estabilizadores coinciden}, the code $\orbbeta(\cF)$ is disjoint if, and only if, for every $1\leq i\leq r$, it holds $\stabbeta(\cF_i)=\stabbeta(\cF)$. Since 
	$\stabbeta(\cF_i)=\langle\beta\rangle\cap\bbF_{q^{m_i}}^\ast$, for every $1\leq i\leq r$, and $\stabbeta(\cF)=\langle\beta\rangle\cap\bbF_{q^m}^\ast$, the result follows. In the particular case of $\beta$ primitive, then it must hold $\stab(\cF_i)=\bbF_{q^m}^\ast$, i.e., the best friend of each $\cF_i$ coincides with the one of $\cF$. 
	\end{proof}

	Observe that it is possible to give a tighter lower bound for the distance of disjoint $\beta$-cyclic orbit flag codes with $\bbF_q^m$ as best friend. In order to avoid codes with distance equal to zero, throughout the rest of the section we only consider elements $\beta\in\bbF_{q^n}^\ast\setminus\bbF_{q^m}^\ast$.
	\begin{proposition}\label{prop: distancia disjunto best friend}
		Let $\cF=(\cF_1, \ldots, \cF_r)$ be a flag on $\bbF_{q^n}$ with the subfield $\bbF_{q^m}$ as its best friend and $\beta\in\bbF_{q^n}^\ast$. If the code $\orbbeta(\cF)$ is disjoint, then $2mr \leq d_f(\orbbeta(\cF))$.
	\end{proposition}
	
	\begin{proof}
		Let $\cF'$ be a flag in $\orbbeta(\cF)$ with $\cF'\neq \cF$. As $\orbbeta(\cF)$ is a disjoint flag code, we have that $\cF_i\neq \cF'_i$ for every $1\leq i\leq r$. Hence, by means of Proposition \ref{prop: stab+ es best friend}, for every $1\leq i \leq r$, we have that $d_S(\cF_i, \cF'_i) \geq 2m$. We conclude that $d_f(\cF, \cF')\geq 2mr$, for every $\cF'\in \orbbeta(\cF)\setminus\{\cF\}$, and the result holds.
	\end{proof}
	As shown in Proposition \ref{prop: cardinality and best friend}, the cardinality of a  $\beta$-cyclic flag code $\orbbeta(\cF)$ with $\bbF_{q^m}^\ast$ as its best friend is completely determined. Moreover, we know that $\langle\beta\rangle=\langle \alpha^l\rangle$, for the divisor $l$ of $q^n-1$ such that $|\beta|=\frac{q^n-1}{l}$. Similarly, $\bbF_{q^m}^\ast=\langle\alpha^{\frac{q^n-1}{q^m-1}}\rangle$. Moreover, it holds
$$
\stabbeta(\cF) = \langle\alpha^{\lcm(l, \frac{q^n-1}{q^m-1})}\rangle \ \text{and} \  |\stabbeta(\cF)|=\frac{q^n-1}{\lcm(l, \frac{q^n-1}{q^m-1})}.
$$
As a result, 
\begin{equation}\label{eq: cardinality beta-cyclic power of alpha}
|\orbbeta(\cF)| = \frac{\lcm \left( l, \frac{q^n-1}{q^m-1}  \right)}{l} .
\end{equation}
Using this notation, the next result follows.
	
	\begin{theorem}\label{theo: type vector ODFC orbital}
Let $\cF$ be a flag on $\bbF_{q^n}$ with best friend $\bbF_{q^m}$. Take $\beta\in\bbF_{q^n}^\ast$ and write $\langle\beta\rangle=\langle\alpha^l\rangle$ with $l$ a divisor of $q^n-1$. If $\orbbeta(\cF)$ is an optimum distance flag code and $t$ is a dimension in the type vector of $\cF$, then $m$ divides $t$ and it must hold
		$$
		\frac{\lcm(l, \frac{q^n-1}{q^m-1})}{l} \leq
		\left\lbrace
		\begin{array}{cll}
			\left\lfloor\frac{q^n-1}{q^t-1}\right\rfloor      & \text{if} & 2t\leq n, \\
			\left\lfloor\frac{q^n-1}{q^{n-t}-1}\right\rfloor  & \text{if} & 2t > n.
		\end{array}
		\right.
		$$
	\end{theorem}
	\begin{proof}
		Consider a flag $\cF$ on $\bbF_{q^n}$ with the subfield $\bbF_{q^m}$ as its best friend and assume that the code $\orbbeta(\cF)$ is an optimum distance flag code. Hence, by application of Lemma \ref{lem: BF divides dimensions}, $m$ must divide every dimension in the type vector. Moreover, by means of Theorem \ref{theo: caracterización ODFC}, all the projected codes attain the maximum possible distance for their dimension and $\orbbeta(\cF)$ is disjoint. In other words, the cardinality of every projected code coincides with $|\orbbeta(\cF)|$. In particular, this value has to satisfy the bounds for the cardinality of constant dimension codes of maximum distance given in Section \ref{sec:Preliminaries} for dimensions in the type vector. As a result, if $t$ is a dimension in the type vector, it must hold:
		\begin{enumerate}
			\item If $2t\leq n$, then $|\orbbeta(\cF)| \leq \left\lfloor\frac{q^n-1}{q^t-1}\right\rfloor$ and
			\item if $2t>n$, then  $|\orbbeta(\cF)| \leq \left\lfloor\frac{q^n-1}{q^{n-t}-1}\right\rfloor$.
		\end{enumerate}
 Moreover, assuming $\langle\beta\rangle=\langle\alpha^l\rangle$ for some divisor $l$ of $q^n-1$, by using (\ref{eq: cardinality beta-cyclic power of alpha}), the result holds.
	\end{proof}
	
	\begin{remark}
		Observe that a dimension $t$ satisfies the necessary condition provided in Theorem \ref{theo: type vector ODFC orbital} if, and only if, the dimension $n-t$ does it as well. This is due to the fact that the upper bound for the cardinality of constant dimension codes with maximum distance of dimensions $t$ and $n-t$ of $\bbF_{q^n}$ coincide. Moreover, these upper bounds decrease as dimensions get closer to $n/2$. Hence, central dimensions are allowed for a smaller number elements $\beta\in\bbF_{q^n}^\ast$ than the other ones. In contrast, extreme dimensions, that is, $m$ and $n-m$, are allowed for every subgroup of $\bbF_{q^n}^*$. In fact, when the acting group is $\bbF_{q^n}^\ast$, we can derive the following corollary.
	\end{remark}
	
	\begin{corollary}\label{cor: type odfc cyclic}
		Assume that the cyclic orbit code $\orb(\cF)$ is an optimum distance flag code on $\bbF_{q^n}$ with the subfield $\bbF_{q^m}$ as its best friend. Then one of the following statements holds:
		\begin{enumerate}
			\item $\orb(\cF)$ is a constant dimension code of dimension either $m$ or $n-m$.
			\item $\orb(\cF)$ has type vector $(m, n-m)$.
		\end{enumerate}
		In any of the three cases above, the code $\orb(\cF)$ has the largest possible size, that is, $\frac{q^n-1}{q^m-1}$.
	\end{corollary}
	
	\begin{proof}
		This result follows by application of Theorem \ref{theo: caracterización ODFC} when $\beta$ is a primitive element of $\bbF_{q^n}^\ast$. In this case, the cardinality of every projected code is $\frac{q^n-1}{q^{m}-1}$. Moreover, if $t$ is a dimension in the type vector, it has to be a multiple of $m$. Observe that, both $m$ and $n-m$ satisfy the necessary condition given in Theorem \ref{theo: type vector ODFC orbital}. On the other hand, this condition is violated by any other multiple of $m$. Hence, only dimensions $m$ or $n-m$ could appear in the type vector of $\cF$. As a result, optimum distance cyclic orbit flag codes with  $\bbF_{q^m}$ as their best friend could only be constructed for type vectors equal to $(m)$, $(n-m)$ or $(m, n-m)$. For these three type vectors, the cardinality of $\orb(\cF)$, which is also $\frac{q^n-1}{q^{m}-1},$ coincides with the largest possible size of constant dimension codes with maximum distance for both dimensions $m$ and $n-m$. Hence, it is the best size for optimum distance flag codes with any of these type vectors.
	\end{proof}
	
	Apart from the case where the type vector is $(m, n-m)$, we see that optimum distance cyclic orbit flag codes with $\bbF_{q^m}$ as their best friend are actually cyclic orbit (subspace) codes of dimension either $m$ or $n-m$. In case the dimension is $m$, the code $\orb(\cF)$ is, in addition, the $m$-spread $\orb(\bbF_{q^m})$ of $\bbF_{q^n}$.
	
	From Theorem \ref{theo: type vector ODFC orbital} and Corollary \ref{cor: type odfc cyclic}, one can deduce that not every type vector is compatible with attaining the maximum possible distance once we have fixed the best friend of the generating flag of a $\beta$-cyclic orbit flag code. The following examples exhibit this fact. 
	\begin{example}
		Let $\cF$ be a flag on $\bbF_{2^{12}}$ with the subfield $\bbF_{2^2}$ as its best friend. This condition implies that the dimensions in the type vector of $\cF$ must be even integers. Notice that $|\bbF_{2^{12}}^\ast|=2^{12}-1=4095= 273 \cdot 15$ and $\langle\alpha^{15}\rangle$  is the only subgroup of $\bbF_{2^{12}}^\ast$ of order $273$. On the other hand, we have $\bbF_{2^2}^\ast=\langle\alpha^{1365}\rangle$. Since $\lcm(15, 1365)=1365$, we have $|\orbbeta(\cF)|=\frac{1365}{15} = 91,$ for every $\beta\in\langle\alpha^{15}\rangle$. Now, assume that $\orbbeta(\cF)$ is an optimum distance flag code. If we compare its size with the upper bounds for the cardinality of constant dimension codes of $\bbF_{2^{12}}$ with maximum distance, we conclude that the dimension $6$ cannot appear in the type vector of $\orbbeta(\cF)$ since $\frac{2^{12}-1}{2^6-1}=65 < 91.$
		In contrast, dimensions $2, 4, 8$ and $10$ satisfy the necessary condition given in Theorem \ref{theo: type vector ODFC orbital}.
	\end{example}
	
	\begin{example}
		Consider a flag $\cF$ on $\bbF_{q^n}$ with the subfield $\bbF_{q^m}$ as its best friend and let $\alpha$ denote a primitive element of $\bbF_{q^n}$. The tables below illustrate which dimensions are susceptible to appear in the type vector of the optimum distance $\beta$-cyclic orbit flag code generated by $\cF$ for different choices of $\beta$ and specific values of $q, n$ and $m$. 
	    
	    	\begin{table}[H]
			\centering
			\begin{small}
			\begin{tabular}{cccccc}
				\hline    
				$\beta$                &$|\beta|$& $\langle\beta\rangle\cap\bbF_{q^m}^\ast$ & $|\orbbeta(\cF)|$& Allowed dimensions  & Max. distance     \\ \hline
				$\alpha$               &  6560   & $\bbF_{3}^\ast$                     &        3280      & 1, 7               & 4 \\
				$\alpha^2$             &  3280   & $\bbF_{3}^\ast$                     &        1640      & 1, 7               & 4  \\
				$\alpha^4$             &  1640   & $\bbF_{3}^\ast$                     &         820      & 1, 2, 6, 7         &12 \\
				$\alpha^5$             &  1312   & $\bbF_{3}^\ast$                     &         656      & 1, 2, 6, 7         &12 \\
				$\alpha^8$             &   820   & $\bbF_{3}^\ast$                     &         410      & 1, 2, 6, 7         &12 \\
				$\alpha^{10}$          &   656   & $\bbF_{3}^\ast$                     &         328      & 1, 2, 6, 7         &12 \\
				$\alpha^{16}$          &   410   & $\bbF_{3}^\ast$                     &         205      & 1, 2, 3, 5, 6, 7   &24 \\
				$\alpha^{20}$          &   328   & $\bbF_{3}^\ast$                     &         164      & 1, 2, 3, 5, 6, 7   &24 \\
				$\alpha^{32}$          &   205   & $\{1\}$                             &         205      & 1, 2, 3, 5, 6, 7   &24 \\
				$\alpha^{40}$          &   164   & $\bbF_{3}^\ast$                     &          82      & 1, 2, 3, 4, 5, 6, 7  &32 \\
				$\alpha^{41}$          &   160   & $\bbF_{3}^\ast$                     &          80      & 1, 2, 3, 4, 5, 6, 7  &32 \\
				$\alpha^{80}$          &    82   & $\bbF_{3}^\ast$                     &          41      & 1, 2, 3, 4, 5, 6, 7  &32 \\
				$\alpha^{82}$          &    80   & $\bbF_{3}^\ast$                     &          40      & 1, 2, 3, 4, 5, 6, 7  &32 \\
				$\alpha^{160}$         &    41   & $\{1\}$                             &          41      & 1, 2, 3, 4, 5, 6, 7  &32 \\
				$\alpha^{164}$         &    40   & $\bbF_{3}^\ast$                     &          20      & 1, 2, 3, 4, 5, 6, 7  &32 \\
				$\alpha^{205}$         &    32   & $\bbF_{3}^\ast$                     &          16      & 1, 2, 3, 4, 5, 6, 7  &32 \\
				$\alpha^{328}$         &    20   & $\bbF_{3}^\ast$                     &          10      & 1, 2, 3, 4, 5, 6, 7  &32 \\
				$\alpha^{410}$         &    16   & $\bbF_{3}^\ast$                     &           8      & 1, 2, 3, 4, 5, 6, 7  &32 \\
				$\alpha^{656}$         &    10   & $\bbF_{3}^\ast$                     &           5      & 1, 2, 3, 4, 5, 6, 7  &32 \\
				$\alpha^{820}$         &     8   & $\bbF_{3}^\ast$                     &           4      & 1, 2, 3, 4, 5, 6, 7  &32 \\
				$\alpha^{1312}$        &     5   & $\{1\}$                             &           5      & 1, 2, 3, 4, 5, 6, 7  &32 \\
				$\alpha^{1640}$        &     4   & $\bbF_{3}^\ast$                     &           2      & 1, 2, 3, 4, 5, 6, 7  &32 \\
				$\alpha^{3280}$        &     2   & $\bbF_{3}^\ast$                     &           1      & 1, 2, 3, 4, 5, 6, 7  & 0 \\
				$1$                    &     1   & $\{1\}$                             &           1      & 1, 2, 3, 4, 5, 6, 7  & 0 \\ \hline
			\end{tabular}
			\end{small}
			\caption{Values for $q=3$, $n=8$, $m=1$ and all  subgroups of $\bbF_{3^8}^\ast$.}\label{table: q=3, n=8, m=1}
		\end{table}
		
As it occurs when considering Galois $\beta$-cyclic flag codes, in these tables we can see that different subgroups of $\bbF_{q^n}^\ast$ (hence, subgroups with different order) can provide the same $\beta$-cyclic orbit flag code. Furthermore, there are different subgroups providing  in turn different orbits but sharing the set of allowed dimensions and, as a consequence, also sharing the maximum possible value for the distance. For instance, in Table \ref{table: q=2, n=12, m=2}, both subgroups $\langle\alpha^{3}\rangle$ and $\langle\alpha^{9}\rangle$ give us the same orbit. On the other hand, the orbits under the action of $\langle\alpha^{5}\rangle$ and $\langle\alpha^{7}\rangle$ have different cardinality (thus, they are different codes) but their sets of allowed dimensions are equal.

		\begin{table}[H]
			\centering
			\begin{small}
		    \begin{tabular}{cccccc}
				\hline
				$\beta$           & $|\beta|$ & $\langle\beta\rangle\cap\bbF_{q^m}^\ast$ & $|\orbbeta(\cF)|$ & Allowed dimensions  & Max. distance    \\ \hline
			    $\alpha$               & 4095   &     $\bbF_{2^2}^\ast$		      &    1365     & 2, 10             & 8 \\
				$\alpha^3$             & 1365   &     $\bbF_{2^2}^\ast$           &    455      & 2, 10             & 8  \\
				$\alpha^5$             & 819    &     $\bbF_{2^2}^\ast$           &    273      & 2, 4, 8, 10       & 24 \\
				$\alpha^{7}$           & 585    &     $\bbF_{2^2}^\ast$           &    195      & 2, 4, 8, 10       & 24 \\
				$\alpha^{9}$           & 455    &     $\{1\}$                     &    455      & 2, 10             & 8 \\
		    	$\alpha^{13}$          & 315    &     $\bbF_{2^2}^\ast$           &    105      & 2, 4, 8, 10       & 24 \\
				$\alpha^{15}$          & 273    &     $\bbF_{2^2}^\ast$           &    91       & 2, 4, 8, 10       & 24 \\
				$\alpha^{21}$          & 195    &     $\bbF_{2^2}^\ast$           &    65       & 2, 4, 6, 8, 10    & 36 \\
				$\alpha^{35}$          & 117    &     $\bbF_{2^2}^\ast$           &    39       & 2, 4, 6, 8, 10    & 36 \\
				$\alpha^{39}$          & 105    &     $\bbF_{2^2}^\ast$           &    35       & 2, 4, 6, 8, 10    & 36 \\
				$\alpha^{45}$          & 91     &     $\{1\}$                     &    91       & 2, 4, 8, 10       & 24 \\
				$\alpha^{63}$          & 65     &     $\{1\}$                     &    65       & 2, 4, 6, 8, 10    & 36 \\
				$\alpha^{65}$          & 63     &     $\bbF_{2^2}^\ast$           &    21       & 2, 4, 6, 8, 10    & 36 \\
				$\alpha^{91}$          & 45     &     $\bbF_{2^2}^\ast$           &    15       & 2, 4, 6, 8, 10    & 36 \\
				$\alpha^{105}$         & 39     &     $\bbF_{2^2}^\ast$           &    13       & 2, 4, 6, 8, 10    & 36 \\
				$\alpha^{117}$         & 35     &     $\{1\}$                     &    35       & 2, 4, 6, 8, 10    & 36 \\
				$\alpha^{195}$         & 21     &     $\bbF_{2^2}^\ast$           &    7        & 2, 4, 6, 8, 10    & 36 \\
    			$\alpha^{273}$         & 15     &     $\bbF_{2^2}^\ast$           &    5        & 2, 4, 6, 8, 10    & 36 \\
    			$\alpha^{315}$         & 13     &     $\{1\}$                     &    13       & 2, 4, 6, 8, 10    & 36 \\
    			$\alpha^{455}$         & 9      &     $\bbF_{2^2}^\ast$           &    3        & 2, 4, 6, 8, 10    & 36 \\
    			$\alpha^{585}$         & 7      &     $\{1\}$                     &    7        & 2, 4, 6, 8, 10    & 36 \\
				$\alpha^{819}$         & 5      &     $\{1\}$                     &    5        & 2, 4, 6, 8, 10    & 36 \\
    			$\alpha^{1365}$        & 3      &     $\bbF_{2^2}^\ast$           &    1        & 2, 4, 6, 8, 10    & 0 \\
    			$1$        & 1      &     $\{1\}$                     &    1        & 2, 4, 6, 8, 10    & 0 \\ \hline 
			\end{tabular}
			\end{small}
			\caption{Values for $q=2$, $n=12$, $m=2$ and all  subgroups of $\bbF_{2^{12}}^\ast$.}\label{table: q=2, n=12, m=2}
		\end{table}
	\end{example}

	\begin{remark}
		Observe that results \ref{theo: type vector ODFC orbital} and \ref{cor: type odfc cyclic} give us necessary conditions on the type vector for the existence of  optimum distance $\beta$-cyclic orbit flag codes but the problem of constucting them remains open. In Subsection \ref{subsec:Galois codes} we have characterized optimum distance Galois $\beta$-cyclic flag codes and built them by providing a suitable subgroup $\langle\beta\rangle$ of $\bbF_{q^n}^\ast$. Recall that in that case, the allowed dimensions correspond to the divisors appearing in the type vector of the generating Galois flag. Looking at Table \ref{table: q=2, n=12, m=2}, for instance, we can obtain optimum distance Galois $\beta$-cyclic flag codes of types $(2,4)$ and $(2,6)$.
	\end{remark}
	
	Apart from optimum distance cyclic flag codes of Galois type, as far as we know, there are only two constructions of optimum distance flag codes given by the action of a cyclic subgroup of $\bbF_{q^n}$. One of them can be found in \cite[Prop. 2.5]{Kurz20}, where the author, for every prime power $q$, provide a cyclic orbit full flag code on $\bbF_{q^3}$ (hence, of type $(1,2)$) with maximum distance as a matching code obtained from the action of $\bbF_{q^3}^\ast$. The same argument allows us to build optimum distance cyclic orbit flag codes with best friend $\bbF_q$ of type $(1, n-1)$ as matching codes for every $n\geq 3$. On the other hand, in \cite{OrbitODFC}, the authors present a construction of an optimum distance orbit full flag code on $\bbF_{q^{2k}}$ arising from the action of a subgroup of $\GL(2k, q)$ that is a cyclic group generated by the companion matrix of a primitive polynomial of degree $2k$ in $\bbF_q[x]$. Observe that this action can be naturally translated into our scenario by identifying such a companion matrix with a primitive element of $\bbF_{q^{2k}},$ as it was pointed out in \cite[Lemma 21]{TrautManBraunRos2013}.

	\section{Conclusions and future work}
	We have introduced the concept of cyclic orbit flag code as a generalization of cyclic orbit (subspace) code to the flag codes setting. Following the viewpoint of \cite{GLMoTro2015}, we analyze the structure and properties of this family of codes by defining the best friend of a flag. This approach allows us to easily compute the cardinality of the code and to provide bounds for its distance. 
	
	In particular, we explore families of codes attaining the extreme possible values for the distance. For the minimum one, we introduce the family of Galois cyclic flag codes, whose elements present a rich structure of nested spreads compatible with the action of $\bbF_{q^n}^*$ on flags. We also study the subcodes of Galois cyclic flag codes whose structure is also orbital cyclic, the Galois $\beta$-cyclic flag codes, and show that we can improve the distance of such codes by choosing a suitable $\beta$ to attain even the maximum possible one. On the other hand, concerning optimum distance flag codes with a fixed best friend, we have provided a necessary condition on the type vector of orbit flag codes that attain the maximum possible distance and arise also from the action of subgroups of $\bbF_{q^n}^\ast$.
	
	In future work we want to come up with other constructions of $\beta$-cyclic orbit flag codes as well as to study conditions and properties of cyclic obit codes with a prescribed distance not necessarily being the maximum one. Despite the study of union of cyclic and $\beta$-cyclic orbit flag codes has not been addressed in this paper, it would be also interesting to tackle this problem. In addition, we would like to exploit the structure of cyclic orbit flag codes in order to determine efficient decoding algorithms taking advantage of the ones already designed for cyclic orbit (subspace) codes in \cite{TrautManBraunRos2013}.

\end{document}